\documentclass[
 article,
 twocolumn,
 groupedaddress,
 showpacs,
 preprintnumbers,
 amsmath,
 amsthm,
 amssymb,
 aps,
 prx,
 floatfix,
]{revtex4-1}

\usepackage{graphicx}					
\usepackage{verbatim}					
\usepackage{dcolumn}                    
\usepackage[many]{tcolorbox}
\usepackage[ruled,vlined]{algorithm2e}
\usepackage{color}
\usepackage{fontawesome}
\usepackage{adjustbox}

\usepackage{orcidlink}                     

\newcommand{\ds}{\displaystyle}             

\newcommand{\dd}{\mathrm{d}}                
\newcommand{\pd}{\partial}                  

\newcommand{\T}{\mathrm{T}}                 
\newcommand{\M}{\mathrm{M}}                 
\newcommand{\Sh}{\mathrm{S}}                
\newcommand{\R}{\mathrm{R}}                 
\newcommand{\Rf}{\mathrm{Ref}}              
\newcommand{\Rt}{\mathrm{Rot}}              
\newcommand{\idt}{\mathrm{I}_2}             

\newcommand{\J}{\boldsymbol{\mathrm{J}}}   

\DeclareMathOperator{\Tr}{tr}               

\newcommand{\const}{\mathrm{const}}

\newcommand{\K}{\mathcal{K}}                

\newcommand{\z}{\zeta}                      

\begin{document}

\title{Isochronous and period-doubling diagrams for symplectic
maps of the plane}
\author{T.~Zolkin}
\email{zolkin@fnal.gov}
\affiliation{Fermilab, PO Box 500, Batavia, IL 60510-5011}
\author{S.~Nagaitsev}
\affiliation{Brookhaven National Laboratory, Upton, NY 11973}
\affiliation{Old Dominion University, Norfolk, VA 23529}
\author{I.~Morozov\,\orcidlink{0000-0002-1821-7051}}
\affiliation{Synchrotron Radiation Facility "SKIF", Koltsovo 630559, Russia}
\affiliation{Novosibirsk State Technical University, Novosibirsk 630073, Russia}
\author{S.~Kladov}
\affiliation{University of Chicago, Chicago, IL 60637}
\author{Y-K.~Kim}
\affiliation{University of Chicago, Chicago, IL 60637}

\date{\today}

\begin{abstract}
Symplectic mappings of the plane serve as key models for exploring
the fundamental nature of complex behavior in nonlinear systems.
Central to this exploration is the effective visualization of
stability regimes, which enables the interpretation of how systems
evolve under varying conditions.
While the area-preserving quadratic H\'enon map has received
significant theoretical attention, a comprehensive description
of its mixed parameter-space dynamics remain lacking.
This limitation arises from early attempts to reduce the full
two-dimensional phase space to a one-dimensional projection, a
simplification that resulted in the loss of important dynamical
features.
Consequently, there is a clear need for a more thorough understanding
of the underlying qualitative aspects.

This paper aims to address this gap by revisiting the foundational
concepts of reversibility and associated symmetries, first explored
in the early works of G.D. Birkhoff.
We extend the original framework proposed by H\'enon by adding a
period-doubling diagram to his isochronous diagram, which allows to
represents the system's bifurcations and the groups of symmetric
periodic orbits that emerge in typical bifurcations of the fixed
point.
A qualitative and quantitative explanation of the main features of
the region of parameters with bounded motion is provided, along
with the application of this technique to other symplectic mappings,
including cases of multiple reversibility.
Modern chaos indicators, such as the Reversibility Error Method (REM)
and the Generalized Alignment Index (GALI), are employed to distinguish
between various dynamical regimes in the mixed space of variables and
parameters.
These tools prove effective in differentiating regular and chaotic
dynamics, as well as in identifying twistless orbits and their
associated bifurcations.
Additionally, we discuss the application of these methods to real-world
problems, such as visualizing dynamic aperture in accelerator physics,
where our findings have direct relevance.
\end{abstract}

\maketitle

\vspace{-0.3cm}
\section{Introduction}

\vspace{-0.2cm}
The study of dynamical systems often reveals intricate structures
and behaviors, bridging mathematics, physics, and computational
science.
By iterating discrete-time mappings or solving differential
equations, we uncover regimes of motion that can be stable or
unstable, periodic or quasiperiodic, regular or chaotic, or,
let say, converge to an attractor.
Domains corresponding to different regimes of motion for various
dynamical systems are known to present complex shapes in the space
of both phase-space variables and the system's parameters.

A cornerstone of exploring complex dynamical systems lies in the
visualization and classification of stability regimes.
These visual tools, such as bifurcation diagrams and control plots,
form the foundation for understanding transitions and behaviors
within these systems.
By offering a visual language, they allow researchers to interpret
how systems evolve under varying parameters and initial conditions.

Among the most famous examples of such visualizations are the
Mandelbrot-Brooks-Matelski (MBM), Julia, and Fatou sets.
These sets collectively serve as a powerful framework for analyzing
the interplay between stability, chaos, and fractal geometries in
the context of iterated complex functions. 
 
\begin{figure*}[t!]
    \centering
    \includegraphics[width=\linewidth]{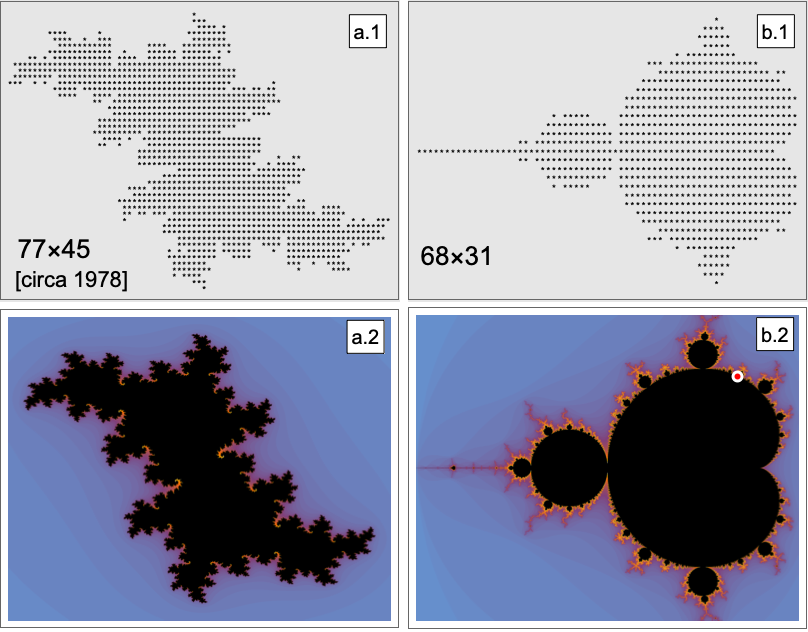}
    \caption{\label{fig:JuliaOG}
    {\bf Julia and MBM sets}.
    The top row shows the Julia set (a.1) for $f(z) = z^2 + C$
    with $C = 0.1+0.6\,i$ and the Mandelbrot-Brooks-Matelski (MBM)
    set (b.1) reproduced from the original work of Brooks and
    Matelski~\cite{BrooksMatelski}.
    The bottom row presents high-resolution, colorized versions of
    these sets, generated using \texttt{Wolfram Mathematica}
    built-in functions.
    The red point in (b.2) corresponds to $0.1+0.6\,i$.
    }\vspace{-0.5cm}
\end{figure*}

The story of these sets begins in the early 1900s with French
mathematicians Gaston Julia~\cite{GastonJulia1918} and Pierre
Fatou~\cite{Fatou:1917A,Fatou:1917B}, who pioneered the study
of iterations of complex functions.
In 1978, Robert W. Brooks and Peter Matelski~\cite{BrooksMatelski}
were the first to depict these structures graphically, with their
original plots reconstructed in pallettes (a.1) and (b.1) of
Fig.~\ref{fig:JuliaOG}.
Shortly after, Benoit Mandelbrot used the computational power
of IBM’s Research Center to create high-resolution visualizations
of the MBM set~\cite{Mandelbrot1980FRACTALAO}, forever linking
these fractal structures to the emerging field of chaos theory.

\newpage
All three sets arise from iterating the quadratic map:
\[
z \mapsto z' = z^2 + C,
\qquad\qquad
z\in\mathbb{C},
\]
with complex parameter
\[
C\in\mathbb{C},
\]
but each set, however, explores a distinct aspect of the system.
The MBM set identifies all complex values of $C$, for which the
iteration, starting from $z_0=0$ remains bounded.
Within the Mandelbrot set, the behavior of the system is stable.
These stable regions correspond to black areas in visualizations,
while points outside the set represent parameters where the orbit
diverges.
These diverging regions are often color-coded to reflect the rate
of escape, see plot (b.2) in Fig.~\ref{fig:JuliaOG}.
The boundary of the Mandelbrot set is a fractal, infinitely complex
and self-similar, serving as a {\it control plot} for mapping
stability.

Each point in the Mandelbrot set corresponds to a unique Julia set,
linking parameter space to phase space, for instance, the red point
in plot (b.2) corresponds to the Julia set depicted in plot (a.2).
Julia sets examine the dynamics for a fixed parameter $C$, focusing
on the phase space of initial conditions $z_0$.
These sets form the boundary between points that generate bounded
orbits and those that escape to infinity.
When $C$ lies inside the Mandelbrot set, the corresponding Julia
set is connected, resembling a cohesive fractal structure.
Conversely, if $C$ is outside the Mandelbrot set, the Julia set is
a Cantor set, consisting of disconnected points scattered across the
plane.
The boundary of the Julia set is where chaotic dynamics dominate,
with points neither escaping nor converging but instead forming
fractal patterns.
Fatou set complements this picture, representing regions of phase
space where the dynamics are smooth and predictable, such as
convergence to fixed points or periodic orbits.

\newpage
Real plane mappings provide two other prominent examples of
quadratic iterations: the dissipative H\'enon–Pomeau
map~\cite{henon1976}, characterized by its strange attractor and
two parameters, and the area-preserving quadratic H\'enon
map~\cite{henon1969numerical}, which depends on a single parameter.
Both are among the most extensively studied dynamical systems
of the plane.

We speculate that the H\'enon–Pomeau attractor has garnered
greater attention due to its broader range of applications.
Notably, in 1993, a colored version of the control plot
illustrating the structure of its parameter space was published
in~\cite{GallasHenon}.
In contrast, the area-preserving H\'enon map has received
significant theoretical attention
\cite{sterling1999homoclinic,dulin2000henon,dullin2000twistless},
with its phase space diagrams setting the standard for
illustrating chaotic dynamics.
Yet, despite its theoretical prominence, comprehensive
visualizations of its mixed variable-parameter space remain
limited, even though this was originally attempted in H\'enon's
seminal work.

This set was intended to be a control plot, summarizing the
continuum of two-dimensional phase space portraits into a
two-dimensional representation of the mapping's parameter
and a coordinate along the symmetry line,
Fig.~\ref{fig:FractalOG}.
However, much like the original image generated by Brooks and
Matelski, which was limited to under 100 pixels in each dimension
(approximately 1 kilopixel), the resolution of the H\'enon set
--- restricted to $199\times159$ pixels (about 30,000 sample
trajectories) --- lacked the clarity needed to capture the full
intricacy of the system's dynamics.

Modern computational tools, such as the built-in functions
\texttt{MandelbrotSetPlot[]} and \texttt{JuliaSetPlot[]}
introduced in \texttt{Wolfram Mathematica 10.0}~\cite{Mathematica},
provide efficient visualization of the MBM set, enabling the
creation of high-resolution, detailed images in seconds
(bottom row in Fig.~\ref{fig:JuliaOG}).
However, similar advances in visualizing the combined
parameter-variable spaces of the area-preserving H\'enon map have
not been fully realized.
As a result, the intricate dynamics of this mapping remain
partially obscured, underscoring the need for further exploration
and development of visualization techniques tailored to such
systems.

More importantly, beyond its low resolution, there is a fundamental
qualitative issue: the dramatic reduction of a 2D phase space to a
1D representation of initial conditions led to the loss of crucial
information, a limitation that H\'enon himself observed.
This gap in understanding highlights the need for a renewed
exploration of these structures, inspired by the numerous classical
studies.

This article aims to address several interconnected goals:

\vspace{0.05cm}
\noindent
{\bf (i)}
Our investigation began with an effort to identify missing
features in H\'enon's diagram, which led us to explore the
concepts of reversibility, associated symmetries, and their
implications for invariant sets.
These ideas, though not completely recognized by H\'enon at
the time, had been anticipated by G.D. Birkhoff during the
same period as Fatou and Julia's pioneering research (please
see~\cite{lewis1961reversible} for the list of references).
In the 1950s, Ren\'e J. DeVogelaere~\cite{devogelaere1950}
further developed this understanding, and later, E.~McMillan
utilized these symmetries to discover his integrable systems
\cite{mcmillan1971problem}.
Despite these advances, as noted in J.A.G.~Roberts' and
G.R.W.~Quispel's comprehensive review~\cite{roberts1992revers},
awareness of these concepts has remained limited across
different scientific disciplines.
In Subsections~\ref{sec:Revers} and \ref{sec:InvariantSet},
we revisit and consolidate key properties of these symmetries.
Further, in Subsections~\ref{sec:SymmetricGroups} and
\ref{sec:Islands}, we discuss how these symmetries extend
beyond isolated periodic orbits to groups of such orbits
that emerge in typical bifurcations.
In doing so, we propose extending the original {\it isochronous
diagram} introduced by H\'enon, which captures bifurcations of
fixed points, to include additional {\it period-doubling diagram}
for 2-cycles and, where applicable, further diagrams for cases
of multiple reversibility.
This collection of two-dimensional plots significantly reduces
the complexity of analyzing the original three-dimensional space
of two variables and a parameter, while effectively revealing
the locations of previously overlooked symmetric periodic orbits.

\vspace{0.05cm}
\noindent
{\bf (ii)} In Section~\ref{sec:Color}, we examine modern
visualization techniques to study dynamical
regimes~\cite{PhysRevE.107.064209, Das_2017}.
These include mode-locking, Frequency Map Analysis
(FMA)~\cite{Laskar1999,laskar2003frequencymapanalysisquasiperiodic},
the Reversibility Error Method (REM)~\cite{PANICHI201653,10.1093/mnras/stx374},
and the Generalized Alignment Index (GALI)~\cite{SKOKOS200730,Skokos2016}.
We demonstrate that REM and GALI are not only effective tools
for distinguishing between regular and chaotic dynamics in phase
space but also highly efficient at detecting bifurcations of
isolated periodic trajectories in parameter space.
Furthermore, these methods can pinpoint twistless bifurcations,
which are associated with the emergence of non-isolated structures
in phase space.

\vspace{0.05cm}
\noindent
{\bf (iii)} Section~\ref{sec:Understanding} provides a qualitative
description of stability diagrams using models such as the chaotic
Arnold circle map and the integrable McMillan mappings.
For a more comprehensive treatment, an auxiliary
article~\cite{zolkin2024MCdynamicsIII} examines the intricate
relationships between McMillan integrable systems and typical
chaotic mappings in standard form.
Using the twist variable, which measures the derivative of the
rotation number with respect to the action variable, we offer a
quantitative description of small-amplitude dynamics and
investigate some numerical aspects of resonance overlap at
larger amplitudes.
A dedicated Subsection~\ref{sec:twistless} explores the twistless
orbit and the bifurcations it induces, including its impact
on Arnold tongues.

\vspace{0.05cm}
\noindent
{\bf (iv)} Finally, in Section~\ref{sec:Results}, we provide
diagrams for various homogeneous power function H\'enon mappings
that are directly linked to horizontal dynamics in model accelerator
lattices, incorporating thin nonlinear magnets such as sextupoles,
octupoles, decapoles, and duodecapoles.
Additionally, we analyze the
Chirikov~\cite{chirikov1969research,chirikov1979universal}
mapping, which models longitudinal motion in accelerator rings with
thin RF stations.
This section bridges theoretical concepts with practical applications,
underscoring the relevance of our findings.


\newpage
\section{\label{sec:HenonSet}H\'enon set}

In H\'enon's original article~\cite{henon1969numerical}, he
considered an area-preserving (symplectic) mapping of the plane
($\mathbb{R}^2\mapsto\mathbb{R}^2$) defined as follows:
\[
\T_2:\qquad
\begin{array}{l}
x' = g_2(x,y),    \\[0.25cm]
y' = h_2(x,y),
\end{array}
\]
where the $(')$ symbol denotes the application of the map, and
$g_2$ and $h_2$ are second-degree polynomials:
\[
\begin{array}{r}
g_2(x,y) = a_1\,x + b_1\,y + c_1\,x^2 + d_1\,x\,y + e_1\,y^2,    \\[0.25cm]
h_2(x,y) = a_2\,x + b_2\,y + c_2\,x^2 + d_2\,x\,y + e_2\,y^2.
\end{array}
\]

Assuming a stable invariant point at the origin:
\[
\big|\Tr\,\dd\T_2(x,y=0)\big| < 2
\]
and using the fact that for an area-preserving map, the Jacobian
determinant must be equal to one:
\[
\J = \det \dd\T_2 =
\begin{vmatrix}
    \pd x'/\pd x & \pd x'/\pd y     \\[0.25cm]
    \pd y'/\pd x & \pd y'/\pd y
\end{vmatrix} = 1,
\]
H\'enon was able to show that by a linear change of coordinates,
the transformation can be reduced to a much simpler form:
\begin{equation}
\label{math:T2Henon}
\T:\qquad
\begin{array}{r}
\ds x' = x\,\cos\psi   - \left[y - F(x)\right]\,\sin\psi,    \\[0.25cm]
\ds y' = x\,\sin\psi\, + \left[y - F(x)\right]  \cos\psi,
\end{array}    
\end{equation}
with only one intrinsic parameter $\psi$ remaining, and quadratic
{\it force function} $F(x)=x^2$.
This parameter is irreducible and represents the rotation angle
in the infinitesimal vicinity of the origin, providing a rotation
number of the fixed point:
\[
    \nu_0 = \frac{\psi}{2\,\pi},
\]
which is independent of the mapping's representation.

He discussed some properties of the mapping, including certain
invariant points: {\it fixed points} ($n=1$) and low order
{\it $n$-cycles} ($n=2,3,4$), defined as
\[
    (x,y) = \T^n (x,y),\qquad\qquad n\in\mathbb{Z}.
\]
For the readers' convenience, we provide all analytical
expressions relevant to the discussion of periodic orbits and
their domains of stability in Appendix {\bf A} at the end of
this article.
Next, he acknowledged that mapping (\ref{math:T2Henon}) and
its inverse can be seen as a convolution of two simpler
area-preserving transformations:
\[
    \T = \Rt(\psi) \circ \Sh_{-F},
    \qquad\qquad
    \T^{-1} = \Sh_{F} \circ \Rt(-\psi),
\]
where $\Rt(\psi)$ is the rotation about the origin by an angle
$\psi$:
\[
\Rt(\psi)=\Rt^{-1}(-\psi):\quad
\begin{bmatrix}
      x'    \\[0.25cm]
      y'
\end{bmatrix} =
\begin{bmatrix}
      \cos\psi &-\sin\psi    \\[0.25cm]
      \sin\psi & \cos\psi
\end{bmatrix}
\begin{bmatrix}
      x    \\[0.25cm]
      y
\end{bmatrix},
\]

\begin{figure*}[p!]
    \centering
    \includegraphics[width=\linewidth]{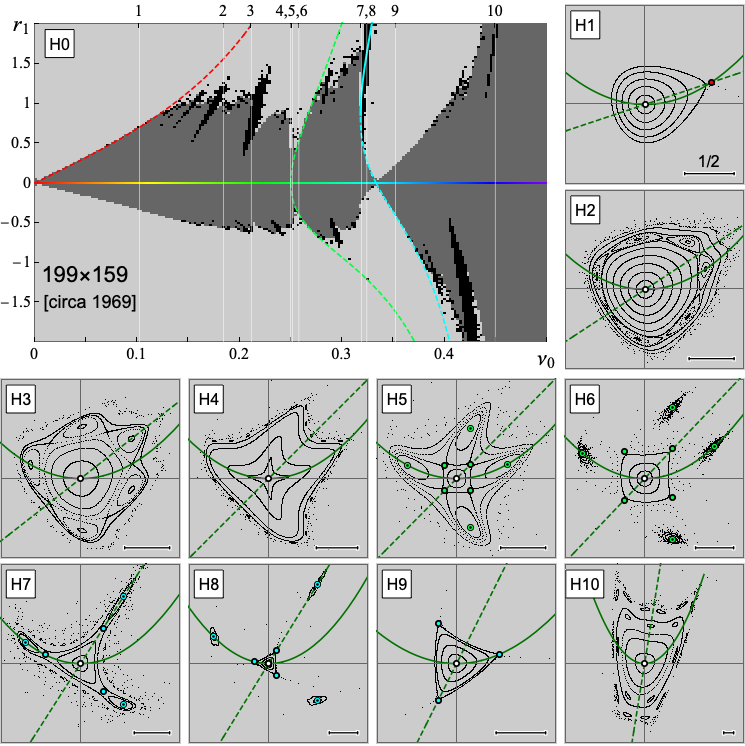}
    \caption{\label{fig:FractalOG}
    {\bf H\'enon set}.
    The main plot at the top [H0] serves as a summary for all values
    of $\nu_0$, depicting distance $r_1$ to initial points along the
    first symmetry line $L_1$ (after H\'enon~\cite{henon1969numerical}).
    Pixels in different colors correspond to various trajectory types:
    light gray indicates escape, dark gray represents successive points
    on a closed curve, and black denotes points on a string of islands.
    Full and dashed curves depict stable and unstable $n$-cycles for
    $n=1,\ldots,4$ respectively, with color indicating the rotation
    number calibrated by $\nu_0$.
    Vertical marks and letters above refer to corresponding phase-space
    portraits $(x,y)$, [H1]--[H10], each displaying several sample
    trajectories (black dots), isolated $n$-cycles (same color as in the
    summary plot), and both symmetry lines (dashed and solid curves in
    dark green for $L_1$ and $L_2$ respectively).
    }
\end{figure*}

\newpage
\noindent
and $\Sh_{\pm F}$ is a nonlinear vertical shear:
\[
\Sh_{\pm F}=\Sh^{-1}_{\mp F}:\qquad
\begin{array}{ll}
    x' = x,                 \\[0.25cm]
    y' = y \pm F(x).
\end{array}
\]
He further remarked on the existence of the symmetry line
\[
L_1:\qquad y = x\,\tan[\psi/2]
\]
which explains the bilateral (mirror) symmetry in phase-space
portraits (refer to the dashed green line in Fig.~\ref{fig:FractalOG})
and is associated with the form of the map, but not the specific
form of the force function $F(x)$.
For illustrations we reconstructed and slightly updated to match
our notation his original plots and combined them in
Fig.~\ref{fig:FractalOG};
as some of his plots were hand drawn, and, in order to incorporate
some additional information minor stylistic changes were made,
while attempting to preserve its original spirit.

He then analyzed long-term stability through numerical
experiments/tracking, utilizing several hundred iterates for each
orbit in specific case studies across various values of the mapping
parameter $\psi$ (plots [H1]--[H10] in Fig.~\ref{fig:FractalOG}).
Moreover, he attempted to capture the entirety of dynamical regimes
in a 2D symbol plot (the main plot [H0] at the top left), where orbit
types were described as a function of the initial position along
the symmetry line $L_1$ and the value of $\nu_0$.
He distinguished three types of trajectories: those that orbit the
origin on a closed curve (stable), those that follow closed curves
around chains of islands (mode-locked), and those that disperse to
infinity (unstable).
However, as mentioned in~\cite{henon1969numerical}, certain
important details, such as even chains of islands, were sometimes
missed, indicating an incompleteness in the summarized depiction
compared to the full 2D phase-space dynamics:

\vspace{0.2cm}
\noindent
{\it ...Plot {\rm [H3]} exhibits a string of 5 large islands.
One of the islands is situated on the positive $L_1$ axis;
this explains the large band on plot {\rm [H0]}.
The same plot {\rm [H0]} shows that these islands move away from the
center and grow larger when $\nu_0$ increases;
they are first inside the curve region (this is the case for plot
{\rm [H3]}), then outside.
Similar bands appear on plot {\rm [H0]} for smaller values of $\nu_0$;
they correspond to strings of 7 and 9 islands.
The chain of 6 islands of plot {\rm [H2]} does not appear on plot
{\rm [H0]} because none of the islands is on the axis of symmetry.
(Incidentally, it seems that all strings with an even number of
islands possess this property, namely, the centers of the islands
are never on the axis of symmetry.
We have not been able to find an explanation for this observed
fact.)...}
\vspace{0.2cm}

In the following, we explore the mystery surrounding H\'enon's set
--- a question partially addressed by G.D. Birkhoff --- and examine
additional methods to enrich the original fractal structure, making
it even more informative.

\newpage
\section{Missing islands}

\subsection{\label{sec:FormOfMap}Form of the map}

Before addressing the issue of missing islands, we may benefit
from making a few key adjustments.
While the form of map~(\ref{math:T2Henon}) is already quite
straightforward, we introduce an alternative representation
referred to here as the McMillan form (see,
e.g.,~\cite{mcmillan1971problem,turaev2002polynomial,ZKN2023PolI,
ZKN2024PolII} for details):
\begin{equation}
\label{math:MHform}
\M:\qquad
\begin{array}{l}
    q' = p,                         \\[0.25cm]
    p' =-q + f(p).
\end{array}
\end{equation}
The two forms of the map are related by a change of variables:
\begin{equation}
\label{math:HMH}
\begin{array}{ll}
    x = p,                          &\qquad
    q = x\,\cos\psi + y\,\sin\psi,  \\[0.25cm]
    y = q\,\csc\psi - p\,\cot\psi,  &\qquad
    p = x,
\end{array}
\end{equation}
with a modified force function:
\[
    f(x) = 2\,x\,\cos\psi + F(x)\,\sin\psi.
\]
While the determinants of these transformations~(\ref{math:HMH})
are constant and independent of dynamical variables, they differ
from unity, introducing additional scaling:
\begin{equation}
\label{math:scaling}
    -\csc\psi\qquad\text{and}\qquad-\sin\psi.
\end{equation}

\noindent
{\bf (i)} H\'enon's consideration of general second-degree
polynomials was constrained by the requirement for a stable fixed
point at the origin, ruling out dynamics beyond the period-doubling
bifurcation.
This could be addressed by using a hyperbolic rotation instead of
the standard rotation matrix, though this requires a further
modification of the map's linear ``normal'' form.
Parametrizing the new force function as
\[
    f(q) = a\,q + q^2,
\]
the McMillan form allows a trace with an absolute value greater
than 2, extending the range of possible dynamical regimes.

\noindent
{\bf (ii)} The H\'enon set extends along $L_1$ when $\nu_0$
approaches a half-integer (1/2), growing infinitely.
The scaling from Eq.~(\ref{math:scaling}) allows us to adjust the
fractal size, making it easier to examine large-amplitude details
in this parameter range.
Subsection~\ref{sec:twisted} discusses the implications of this
scaling in greater detail.

\noindent
{\bf (iii)} The McMillan form of the map can also be decomposed
as a composition of two simpler symplectic mappings:
\[
\M = \Sh_f\circ\Rt(\pi/2),
\qquad
\M^{-1} = \Rt(-\pi/2)\circ\Sh_{-f}.
\]
Additionally, as noted by E. McMillan~\cite{mcmillan1971problem},
there is another decomposition:
\[
\M = \R_f\circ\Rf(\pi/4),
\qquad\quad\,
\M^{-1} = \Rf(\pi/4)\circ\R_f,
\]
where $\mathrm{Ref(\psi)}$ represents a reflection about a line
through the origin at angle $\psi$ with the $q$ axis:
\[
\mathrm{Ref}(\psi):\quad
\begin{bmatrix}
    q'\\[0.25cm] p'
\end{bmatrix} =
\begin{bmatrix}
    \cos2\psi & \sin2\psi\\[0.25cm]
    \sin2\psi &-\cos2\psi
\end{bmatrix}
\begin{bmatrix}
    q \\[0.25cm] p
\end{bmatrix}
\]
and $\R_f$ is a nonlinear vertical reflection:
\[
\R_f:\qquad
\begin{array}{ll}
    q' = q,                 \\[0.25cm]
    p' =-p + f(q).
\end{array}
\]
Both reflections ($\R$) have a Jacobian determinant of $-1$,
making them involutory transformations with
\begin{equation}
\label{math:InvTr}
    \R^{-1} = \R,
    \qquad
    \R^2 = \idt,
\end{equation}
where $\idt$ is the $2 \times 2$ identity matrix.
As a result, each reflection has a line of fixed points:
\[
l_1:\,\mathrm{Fix}\,[\Rf(\pi/4)] = \left\{ (q,p) |\,p=q \right\}
\]
and
\[
l_2:\,\mathrm{Fix}\,[\R_f] = \left\{ (q,p) |\,p=f(q)/2 \right\},
\]
with all other initial conditions forming 2-cycles under $\R_{1,2}$,
according to Eq.~(\ref{math:InvTr}).
Thus, this decomposition reveals two symmetry lines.

For the transformation in its original form, we have:
\[
\T = \Rf(\psi/2)\circ\R_F,
\qquad\quad\,
\T^{-1} = \R_F\circ\Rf(\psi/2),
\]
with corresponding first and second symmetry lines:
\[
L_1:\,y=x\,\tan(\psi/2)
\qquad\text{and}\qquad
L_2:\,y=F(x)/2.
\]
Comparing these two forms of the map, we see that in the McMillan
form, the first symmetry line $l_1$ is fixed and independent of
$f(q)$, while in H\'enon's form, the first symmetry line $L_1$
rotates as $\nu_0$ increases, and the second symmetry $L_2$
is determined solely by $F(q)$, which is independent of $\psi$.

Another modification we adopt in the McMillan form, unlike
H\'enon's original setup, is to use the horizontal projection
onto the $q$-axis, denoted $q_{1,2}$ for $L_{1,2}$, instead of
measuring distance along the symmetry line $q_1^\dagger$.
This projection is simpler than calculating distance, especially
for the second symmetry, which is defined by a curve rather than
a line.
In the H\'enon form, however, using this projection is less
convenient, as the orientation of $L_1$ depends on $\psi$.

Long before McMillan's work, G.D. Birkhoff and, later,
R.~deVogelaere~\cite{devogelaere1950} had already explored the
decomposition of transformations into two involutions,
highlighting the significance of reversibility in a series of
publications from 1914 to 1945 (see~\cite{lewis1961reversible}
and references therein).
In the following section, we review the fundamental properties
of reversibility and consider its implications, laying the
groundwork for a deeper understanding.

\newpage
\subsection{\label{sec:Revers}Symmetry lines and reversibility}

Despite the long history of time-reversibility, A.J. Roberts
observed in his 1992 report~\cite{roberts1992revers} that
``{\it...reversible dynamical systems have received far less
attention and the treatment that they have received has tended
to be less systematic.
This has led to the situation where the literature on the subject
is quite scattered, with some authors not being aware of the
generality of the mathematical theory underlying their results.}''
Here, we summarize relevant findings
from~\cite{lewis1961reversible,devogelaere1950}, as well
as~\cite{roberts1992revers}, to provide context for further
discussion.

A transformation $\T$ is termed {\it reversible}, if it can be
expressed as a composition of two involutions:
\begin{equation}
\label{math:revers}    
    \T = \R_2\circ\R_1,
    \qquad\qquad
    \R_{1,2}^2 = \idt,
\end{equation}
where $\R_1$ and $\R_2$ are the {\it reversing symmetries} of $\T$.
As discussed in detail in~\cite{roberts1992revers}, this definition
of reversibility does not require $\T$ to be conservative,
area-preserving, measure-preserving, or even to operate on an
even-dimensional manifold.
For a reversible map~(\ref{math:revers}), we find:
\[
    \R_1 = \R_2\circ\T =\T^{-1}\circ\R_2,
    \quad\,\,
    \R_2 = \T\circ\R_1 =\R_1\circ\T^{-1},
\]
which ensures that $\T$ is invertible, with
\[
    \T^{-1} = \R_1\circ\R_2.
\]

This implies that the map and its inverse are {\it conjugate} to
each other, as there exists an invertible {\it conjugating
transformation} $\mathrm{P}$ such that
\begin{equation}
\label{math:conjug}
    \T^{-1} = \mathrm{P}\circ\T\circ\mathrm{P}^{-1},
\end{equation}
because, for both symmetries, we have
\[
\T^{-1} = \R_1\circ\T\circ\R_1
\quad\text{and}\quad
\T^{-1} = \R_2\circ\T\circ\R_2.
\]
Mappings that satisfy~(\ref{math:conjug}) with $\mathrm{P}$
not necessarily an involution are considered {\it weakly
reversible}.

As deVogelaere noted~\cite{devogelaere1950}, if $\R$ is a
reversing symmetry of $\T$, then so is the entire family of
symmetries $\left\{\T^k\circ\R\right\}$, forming an infinite
group along with the iterates of the map $\left\{\T^k\right\}$.
A map can also possess additional, independent families of
reversing symmetries, not necessarily but often weakly reversible,
in which case it is called a {\it multiply reversible} map.
A notable case discussed in~\cite{roberts1992revers} is that of
doubly reversible mappings, such as reversible odd maps, where
$\T$ commutes with the rotation $\Rt(\pi)$.

Lastly, the set of fixed points of $\R$, $\mathrm{Fix}\,\R$, and
those of $\mathrm{Fix}\,\T^n\R$, form an infinite family of
symmetry lines.
When $\R$ is an orientation-reversing $C^1$ involution of the plane,
these lines are non-intersecting and do not terminate, remaining
analytic if $\R$ is analytic, as originally discussed in J.D. Finn's
dissertation.
For maps in H\'enon form, we denote the symmetry lines as $L_{1,2}$,
while in McMillan form, we use $l_{1,2}$ for the first two symmetry
lines $\mathrm{Fix},\R_{1,2}$.

\begin{figure*}[t!h]
    \centering
    \includegraphics[width=\linewidth]{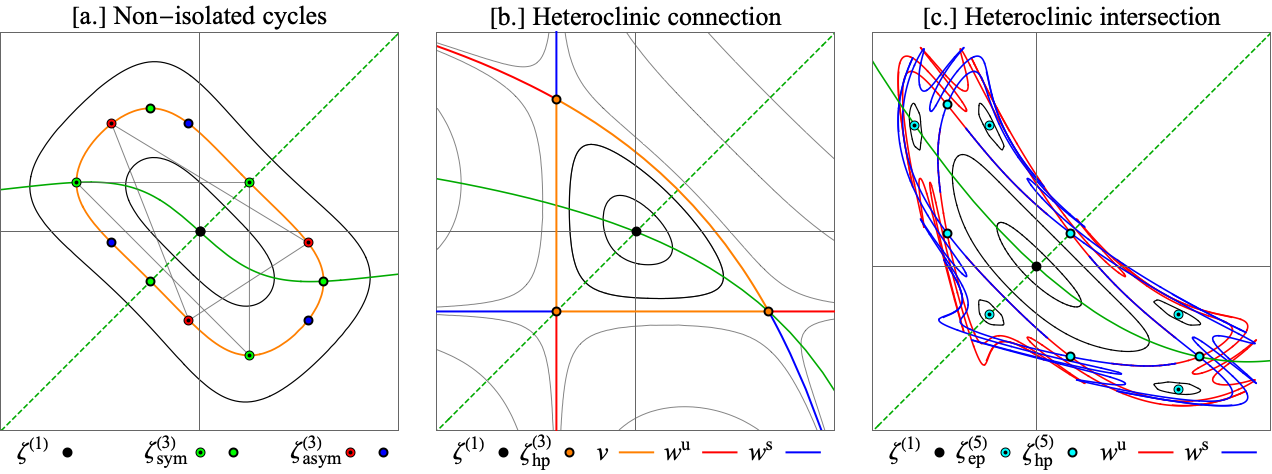}
    \caption{\label{fig:nCycles}
    {\bf Invariant orbits}.
    Plot (a.) demonstrates non-isolated $n$-cycles using the McMillan
    map with $f(q)=-\frac{9}{5}\,q/(q^2+1)$.
    The closed orange curve represents an invariant level set with
    rotation number $1/3$.
    Symmetric orbits are shown in green, while asymmetric orbits
    appear in blue and red.
    Connecting lines illustrate iterations between points.
    Plot (b.) shows a heteroclinic connection in another integrable
    McMillan map, $f(q)=\frac{16}{5}\,q/(q-4)$.
    Red and blue curves depict the unstable ($w^\mathrm{u}$) and
    stable ($w^\mathrm{s}$) manifolds associated with the hyperbolic
    3-cycle $\z^{(3)}_\mathrm{hp}$, with the heteroclinic connection
    $v$ in orange.
    Plot (c.) uses the chaotic H\'enon map $f(q)= -\frac{9}{5}\,q+q^2$,
    revealing a chain of 5 islands with stable and unstable manifolds
    intersecting heteroclinically.
    In all plots, the fixed point at the origin is marked in black,
    while other periodic orbits are colored as per legend.
    Black orbits represent irrational rotations, and gray in (b.)
    indicates unstable trajectories.
    Green dashed and solid lines denote the first, $p=q$, and
    second, $p=f(q)/2$, symmetry lines.
    }
\end{figure*}

\subsection{\label{sec:InvariantSet}Invariant sets}

In this subsection, we define and review some fundamental
properties of invariant sets and orbits in area-preserving
mappings, following~\cite{roberts1992revers}.
Fig.~\ref{fig:nCycles} illustrates these concepts with phase
space portraits of various integrable and chaotic systems.

For a stable orbit of a point $\z_0=(q_0,p_0)$ under an
area-preserving mapping $\T$, the trajectory typically exhibits
one of three behaviors in phase space:

\vspace{0.3cm}
\noindent
{\bf (i)} The trajectory forms a zero-dimensional set of $n$
distinct points visited in a unique periodic sequence:
\begin{equation}
\label{math:ncycle}
    \T^n\z_0 = \z_0,
    \qquad\qquad
    n > 0.
\end{equation}
This set of points is called an $n$-{\it cycle}:
\[
    \z^{(n)} = \left\{\z_0,\z_1,\ldots,\z_{n-1}\right\},
\]
or a {\it fixed point} if $n=1$.
In Fig.~\ref{fig:nCycles} the fixed point at the origin is shown
in black, while other periodic orbits ($n>1$) are color-coded.

\vspace{0.3cm}
\noindent
{\bf (ii)} The trajectory forms a one-dimensional set that lies
on an {\it invariant curve}, $C$, in the plane.
In chaotic systems, these curves correspond to {\it KAM curves}
or {\it circles} and are densely filled in a quasiperiodic manner.
The rotation number for such curves is irrational, resulting in
an angular increment that is incommensurate with $2\,\pi$
(black closed curves in all plots).
In nonlinear integrable systems, however, the rotation number is
often a continuous function of amplitude.
Consequently, when $\nu$ crosses rational values $m/n$, such
invariant curves break into an infinite number of $n$-cycles,
as shown by the orange curve in plot (a.).
We refer to these as {\it non-isolated} $n$-cycles, in contrast
to {\it isolated} $n$-cycles, which lack other orbits of the same
period nearby (provided their eigenvalues are not roots of unity).
In chaotic systems, rational rotation numbers typically correspond
to centers $\z^{(n)}_\mathrm{ep}$ (elliptic periodic points)
or nodes $\z^{(n)}_\mathrm{hp}$ (hyperbolic periodic points)
of island chains.

\vspace{0.3cm}
\noindent
{\bf (iii)} The trajectory appears to wander without period or
quasiperiodicity, densely covering a region of the phase space.
In these cases, the orbit exhibits exponential sensitivity to
initial conditions and is classified as {\it chaotic}.

A set of points, $\Gamma$, is called {\it invariant} under
the mapping $\T$, if $\T\,\Gamma = \Gamma$.
Thus, by definition (\ref{math:ncycle}), all points in an
$n$-cycle form an invariant set.
Likewise, a curve $C$ on the plane can be invariant if
\[
    \forall\,\z\in C:
    \qquad
    \z' = \T\,\z \in C.
\]
In Fig.~\ref{fig:nCycles}, all closed curves around the fixed
point at the origin are invariant, while a curve surrounding
one of the five islands (in plot c.) is not;
here, each curve is sequentially visited, making all five curves
together an invariant set.

\newpage
Stable (elliptic) periodic points are typically surrounded by
elliptical orbits, hence the name.
In contrast, each point in an unstable (hyperbolic)  $n$-cycle
has associated stable ($w^\mathrm{s}$) and unstable 
($w^\mathrm{u}$) manifolds:
\[
w^\mathrm{s,u}\left[\zeta^{(n)}_k\right] = \left\{
    \z_0\,:\,\lim_{N\rightarrow\infty}
        \T^{\pm (n \cdot N)}\z_0 =
    \zeta^{(n)}_k
\right\}
\]
aligned with the eigenvectors of $\dd\T^{n}[\z^{(n)}_k]$
near $\zeta^{(n)}_k$.
For $n>1$, the orbit hops between branches of the same stability
(similar to transitioning between islands) while following the
manifolds either toward or away from the cycle.
Thus, all stable or all unstable branches collectively form
another example of an invariant set.

In integrable systems, these manifolds may extend to infinity
(as shown by the red and blue curves in plot b.) or form a
{\it connection} (orange curve):
\[
    v = w^\mathrm{s}\left[\z^{(n)}_i\right] \bigcap
        w^\mathrm{u}\left[\z^{(n)}_j\right]
\]
called {\it heteroclinic} if $i \neq j$ or {\it homoclinic} if
the point connects to itself, $i = j$.
In chaotic systems, stable and unstable manifolds may intersect,
forming an intricate, ongoing network of
{\it heteroclinic/homoclinic intersections} --- often called
{\it tangles} (see plot c.).

We note that if $\Gamma$ is an invariant set, then so is
$\R\,\Gamma$, where $\R$ is one of the involutions
of $\T$.
For example, the stable (unstable) manifold of a point $\z^{(n)}_k$
maps under $\R$ to the unstable (stable) manifold of
$\R\,\z^{(n)}_k$.
An invariant set is termed {\it symmetric}, if it is invariant under
both $\T$ and $\R$.
This definition can apply to a curve, a periodic or aperiodic orbit,
or a collection of such objects.

As McMillan observed~\cite{mcmillan1971problem}, for integrable
systems, constant level sets, $\K[p,q]=\const$, must be symmetric:
\[
\K[\z_0] = \K[\T\,\z_0]
\quad\text{and}\quad
\K[\z_0] = \K[\R\,\z_0].
\]
In chaotic systems, these level sets correspond to collections of
orbits sharing an ``energy'' level that persists despite
perturbations.
In plot (a.) of Fig.~\ref{fig:nCycles}, the closed orange curve
with a rotation number of $1/3$ represents a symmetric invariant
set.
Among the trajectories on it, only the two shown in green are
symmetric by themselves.
Other orbits, such as those shown in red and blue, are
{\it asymmetric}.
However, the red and blue orbits are constructed as reflections
of each other with respect to $\R$ and, as a pair, they
form a symmetric invariant set (not an orbit!).

The same principle applies to manifolds $w^\mathrm{u,s}$, which,
while individually invariant only under $\T$, together
form a symmetric set.
However, in this case, the intersection of the two unions,
$\mathrm{V} = \Gamma \bigcap \R\,\Gamma$, can be nonempty,
making $\mathrm{V}$ a symmetric invariant (strict) subset of
$\Gamma$.
Thus, invariant sets generally consist of symmetric invariant
subsets alongside pairs of asymmetric subsets.
For instance, in plot (b.) we observe a symmetric 3-cycle along
with a pair of associated stable and unstable manifolds.
In contrast, plot (c.) presents a scenario where, instead of a
periodic symmetric subset, we have a symmetric orbit that
intersects both manifolds and exhibits perpetual wandering.

\subsection{\label{sec:SymmetricGroups}Symmetric groups}

Since for any point in a symmetric orbit,
$\z\in\Gamma_\mathrm{sym}$,
\[
\exists\,k\in\mathbb{Z}:\,\,\R\,\z =
    \T^k\z\in\Gamma_\mathrm{sym},
\]
its trajectory must ``hop across'' the symmetry line if $k$ is odd,
\begin{equation}
\label{math:hop}
    \T^{-1}\circ\R\,\left( \T^{(k-1)/2}\z \right) =\T^{(k-1)/2}\z
\end{equation}
or ``cross'' it if $k$ is even,
\begin{equation}
\label{math:cross}
    \R\,\left( \T^{k/2}\z \right)= \T^{k/2}\z.
\end{equation}
Conversely, if a trajectory crosses~(\ref{math:cross}) the
symmetry line $\mathrm{Fix}\,\R$, it is symmetric with respect
to $\R$ and the entire family of symmetries $(\T^n\circ\R)$.
This is evident as each point of its orbit lies on the symmetry line
according to
\begin{equation}
\label{math:fix}    
\begin{array}{l}
\T^i\,\mathrm{Fix}\,(\R_1) = \mathrm{Fix}\,(\T^{2\,i}\circ\R_1),
\\[0.25cm]
\T^i\,\mathrm{Fix}\,(\R_2) = \mathrm{Fix}\,(\T^{2\,i+1}\circ\R_1).
\end{array}
\end{equation}

If a point lies at the intersection of two distinct symmetry
lines, it is termed {\it doubly symmetric}.
Each doubly symmetric point is periodic under $\T$, and each
symmetric periodic point is doubly
symmetric~\cite{lewis1961reversible,devogelaere1950},
as indicated by
\begin{equation}
\label{math:doubly}
    (\T^m\circ\R)\circ(\T^n\circ\R)\,\z = \T^{m-n}\,\z = \z.
\end{equation}
Thus, for symmetric periodic orbits, each point lies on an
intersection of the entire subfamily of symmetries.
Eqs.~(\ref{math:fix},\ref{math:doubly}) further imply that if $\z$
and $\T^r\z \neq \z$ lie on the same symmetry line, the orbit has
an even period of $2r$.
Moreover, a symmetric periodic orbit has an even period if and only
if it includes two points on the same symmetry line.

\begin{figure}[h!]
    \centering
    \includegraphics[width=\linewidth]{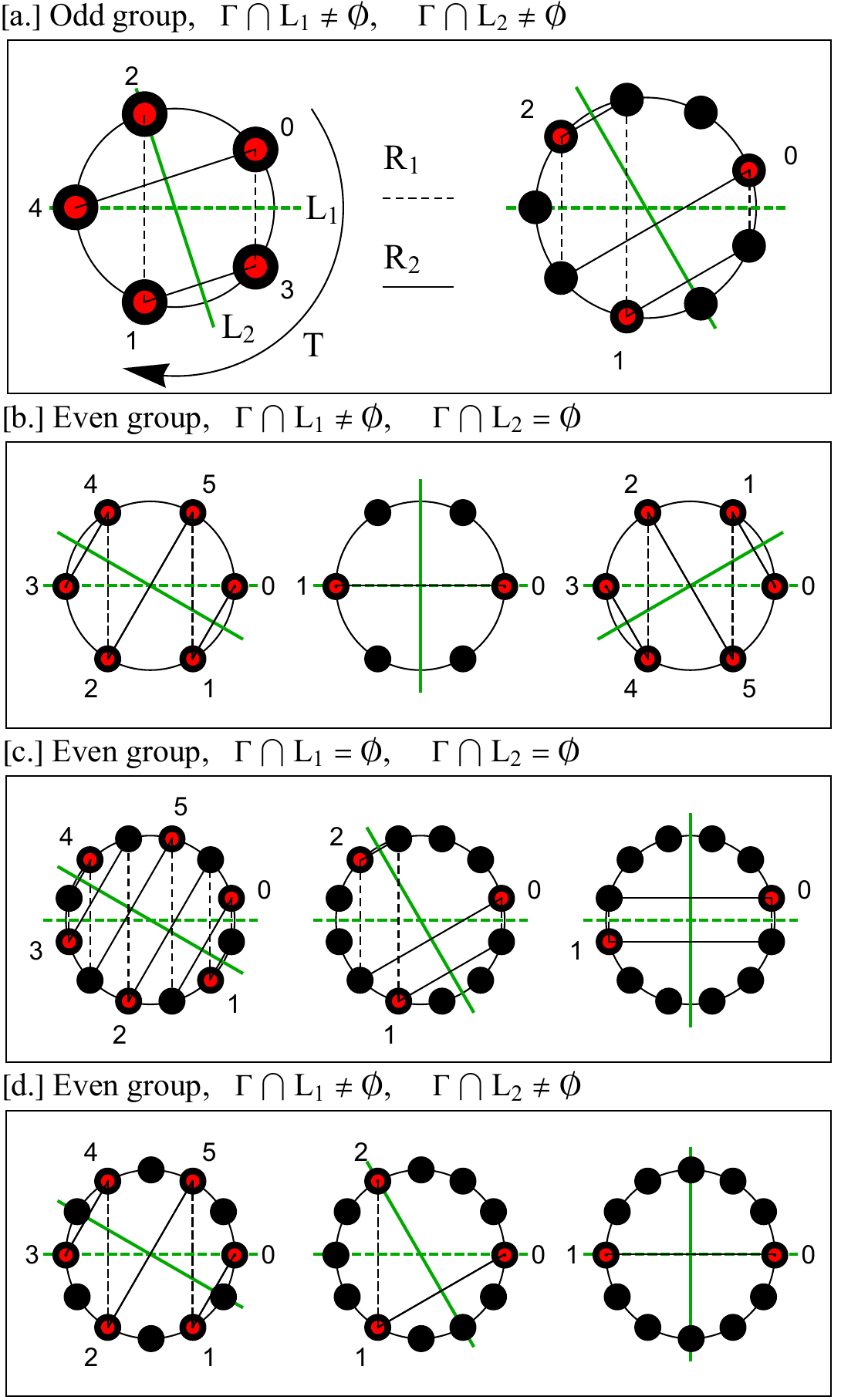}
    \caption{\label{fig:RefRot}
    {\bf Symmetric groups}.
    Plot (a.) illustrates the dynamics of odd symmetric groups with
    $n=5$ and $9$ points.
    The next three plots show the dynamics of even symmetric groups:
    one that intersects only a single symmetry line (b.), one that
    intersects no symmetry lines (c.), and one that intersects both
    (d.).
    In all plots, the first symmetry line $L_1 = \mathrm{Fix}\,\R_1$
    is shown as a dashed green horizontal line, and the second
    symmetry line $L_2 = \mathrm{Fix}\,\R_2$ is solid.
    Dashed and solid black lines represent the iterates of a point
    under the reflections $\R_{1,2}$, respectively.
    Each diagram includes a sample periodic orbit (in red), with
    numerals indicating the iterate number.
    Since the combination of two reflections is equivalent to a
    rotation, the orbits traverse the intersection of the symmetry
    lines in a clockwise direction.\vspace{-0.5cm}
    }
\end{figure}

While much attention has been given to symmetric periodic orbits,
isolated cycles with similar stability can appear in degenerate
groups --- on the same energy level --- alongside asymmetric
orbits.
As we'll see, such groups often split.
To clarify the dynamics of these groups, we consider a symbolic
model illustrated in Fig.~\ref{fig:RefRot}.
In this model, a symmetric group of $n$ points (in black) is
divided evenly by each of the two symmetry lines (in green).
Each iteration, $\T=\R_2\circ\R_1$, is a sequential application
of two involutions, here represented by linear reflections 
$\Rf(\theta)$.
By aligning the first symmetry line with the horizontal axis
($\theta_1=0$), and using
\[
\Rf(\theta_2)\circ\Rf(\theta_1) = \Rt(2\,\theta_2-2\,\theta_1)
\]
we see that the rotation number must satisfy
\[
    \nu = 2\,k/n
\]
where $k$ is the angle between symmetries divided over $2\,\pi$.

For an odd symmetric group or individual  $n$-cycle, geometry
dictates that each symmetry line contains one point of the
group/cycle, as this is the only way to evenly split an odd
group.
These orbits cannot have more than one crossing per line,
implying an even period.
If $\nu$ is a reducible fraction, the group splits into $n/m$
$m$-cycles, where $l/m$ is an irreducible form of $\nu$, as
illustrated in Fig.~\ref{fig:RefRot} (a.) for $n=9$ and $2\,k=3$.

\begin{figure*}[t!]
    \centering
    \includegraphics[width=\linewidth]{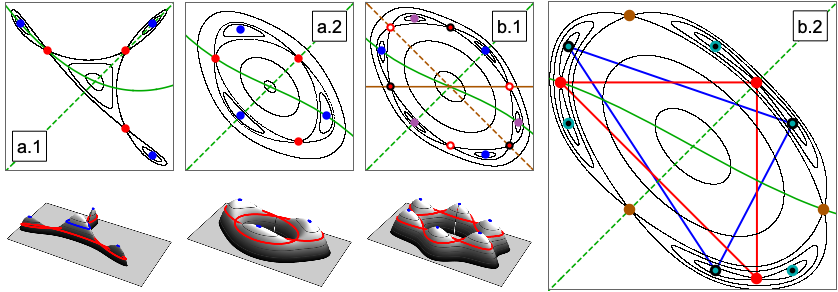}
    \caption{\label{fig:3Cycles}
    {\bf Islands and symmetries}. The top row shows phase-space
    portraits for various H\'enon mappings with force functions
    of the form $f(q) = a\,q + b\,q^2 + c\,q^3$.
    Each plot illustrates the result of a specific bifurcation,
    with the corresponding triplets of coefficients $[a,b,c]$:
    (a.1) saddle-node ($\mathrm{SN}_{1/3}$) $[-0.84, 1, 0]$, or
    a touch-and-go bifurcation if the image is cropped to remove
    the blue points;
    (a.2) 3-island chain $[-0.9, 0.02, -1]$;
    (b.1) doubled 3-island chain $[-0.85, 0, -1]$; and
    (b.2) asymmetric split $[-1.1, 6\times10^{-3}, 1]$.
    The bottom row illustrates the invariant level sets of an
    integrable system with a similar phase-space topology.
    In all plots, isolated 3-cycles are shown in color, and the
    first and second symmetry lines are indicated by solid and
    dashed green lines.
    Plot (b.1) also shows an additional independent family of
    symmetries in brown.
    }
\end{figure*}

For odd symmetric cycles or groups, there is a single crossing
and one hop per symmetry line.
In contrast, even groups are more complex: each symmetry line may
exhibit either two crossings or two hops.
Symmetric even cycles can only cross one symmetry line or must
split, as shown in Fig.~\ref{fig:RefRot} (d.).
Degeneracies occur again if $\nu$ is reducible, as illustrated in
the middle diagram of Fig.~\ref{fig:RefRot} (b.) for $n = 6$ and
$2\,k = 3$.

The two bottom plots explore cases where the even group crosses
neither symmetry line (c.) or both symmetry lines (d.).
In these cases, both $2\,k$ and $n$ must be even, requiring the
group to split, at least in half, consistent with the fact that
symmetric even cycles cannot cross both lines.
Only in case (c.) does the group completely disappear.
In all other cases, even with degeneracies, at least one symmetry
line intersects one of the orbits.
Asymmetric periodic orbits that do not cross symmetry lines appear
in pairs.
Thus, although these orbits do not lie on symmetry lines (case c.),
if they represent centers or nodes of chains of islands, the
connections and intersections of their surrounding manifolds
$w^\mathrm{u,s}$ must form symmetric sets.

\subsection{\label{sec:Islands}Islands and bifurcations}

Finally, we apply the insights we've gathered to understand how
island chains, which emerge from typical bifurcations around fixed
points, interact with the two primary symmetry lines.
While we do not aim to fully classify symplectic map bifurcations,
we follow~\cite{barrio2009} and refer the reader to~\cite{marsden}
for a comprehensive description.

All fixed points and 2-cycles lie on intersections at
$L_1 \cap L_2$ and $L_2 \cap L_2^{-1}$, respectively.
Thus, by analyzing both symmetry lines, we reliably encounter
isochronous bifurcations (i.e., those involving appearance or
disappearance of solutions without changing cycle frequency),
such as transcritical (T), saddle-node (SN), pitchforks (PF),
and period-doubling (PD) bifurcations.
Accordingly, we suggest naming diagrams along $L_1$ as
{\it isochronous} and those along $L_2$ as
{\it period-doubling diagrams}.

Turning to cycles with periods $n\geq 3$, Figure~\ref{fig:3Cycles}
illustrates examples of more complex bifurcations through
phase-space plots, featuring isolated period-3 orbits.

\noindent
{\bf Touch-and-Go (TG)}: In this simple bifurcation, an $n$-cycle
goes through a fixed point without altering its stability.
As the cycle surrounds the fixed point, it must be symmetric,
appearing on symmetry lines according to its parity.
    
\noindent
{\bf Saddle-Node}: Here, a pair of stable and unstable $n$-cycles
emerges from a cusp along a curve (e.g., $\mathrm{SN}_{1/3}$
bifurcation in the quadratic H\'enon map), as shown in
Fig.~\ref{fig:3Cycles} (a.1).
Due to their opposing stability (i.e., distinct energy levels),
each cycle is an invariant set and remains symmetric.
    
\noindent
{\bf $n$-Island Chain}: Illustrated in Fig.~\ref{fig:3Cycles}
(a.2), this case differs from (a.1) in that the manifolds $w^{u,s}$
span across symmetry lines with heteroclinic intersections, rather
than homoclinic ones.
While in the previous case, both the node and the center of the
islands must appear on the same side of the same symmetry line,
now, for odd $n$, stable and unstable solutions appear on opposite
sides of each symmetry line.
For even $n$, cycles appear on both sides of two different
symmetry lines.
Once again, both stable and unstable $n$-cycles form symmetric
orbits independently, with each crossing a symmetry line.

\noindent
{\bf Doubled $n$-island Chain}: Plot (b.1) illustrates this case,
where a group of two unstable 3-cycles (shown in red/white and
red/black) is absent on both symmetry lines, $L_{1,2}$.
This phase plot corresponds to the cubic H\'enon map with an odd
force function $f(q) = a\,q + q^3$, resulting in multiple
reversibilities.
By considering an additional family of independent symmetries
along $L_1^*:\,p=-q$ and $L_2^*:\,p=0$ (see Section oct), we can
locate the missing nodes.

\noindent
{\bf Asymmetric bifurcation}: Finally, plot (b.2) illustrates the
case where the force function is neither even nor odd, resulting
in a pair of asymmetric 3-cycles (in cyan) that fully evade
symmetry lines.
However, as noted earlier, these cycles are surrounded by two
asymmetric homoclinic intersections with a symmetric unstable
3-cycle centered within them.
Consequently, their presence is indirectly observable in our
diagrams, as the stability of a symmetric point will necessarily
shift.
    
\section{\label{sec:Color}Fractal Coloring Methods}

Building on the analysis above, H\'enon's stability diagram can
be refined in several key ways.
First, increasing the number of iterations allows for finer
fractal resolution, capturing more detailed structures.
Second, adding a period-doubling diagram along the second
symmetry line provides conceptual completeness.
Finally, by using color, we can achieve a deeper distinction
between types of trajectories, particularly highlighting
differences between regular and chaotic motion.

H\'enon's original work used phase-space portraits, generated by
tracing a few carefully chosen initial conditions, each revealing
a typical trajectory.
To make these plots representative of the system's behavior,
either deep knowledge of the $n$-cycle stability or a wide
sampling of trajectories is needed.
Despite the limitations, these portraits became widely accepted
due to their simplicity and effectiveness in visualizing dynamical
systems.

Directly assigning colors to each trajectory can result in visual
clutter, as distant initial conditions in phase space may belong
to the same invariant curve.
A more effective approach uses intrinsic properties as color
variables --- e.g., the rotation number for stable trajectories.
With advances in computational power, modern
algorithms~\cite{PhysRevE.107.064209, Das_2017} have enabled
high-resolution scientific visualizations that reveal complex
behaviors in chaotic dynamics.
These methods enable dense sampling across the phase plane, with
points colored by corresponding indicators.
The Frequency Map Analysis
(FMA)~\cite{Laskar1999, laskar2003frequencymapanalysisquasiperiodic},
for instance, tracks the variation in rotation numbers to reveal
areas of its numerical diffusion.
Other techniques, such as the Reversibility Error Method
(REM)~\cite{PANICHI201653, 10.1093/mnras/stx374} or
chaos indicators based on the Lyapunov exponents like the fast
Lyapunov indicators (FLIs)~\cite{FROESCHLE1997881}, along with the
Smaller (SALI) and the Generalized (GALI) Alignment
Indices~\cite{SKOKOS200730, Skokos2016}  proved to be especially
effective at distinguishing chaotic from regular dynamics.

\begin{figure*}[p!]
    \centering
    \includegraphics[width=\linewidth]{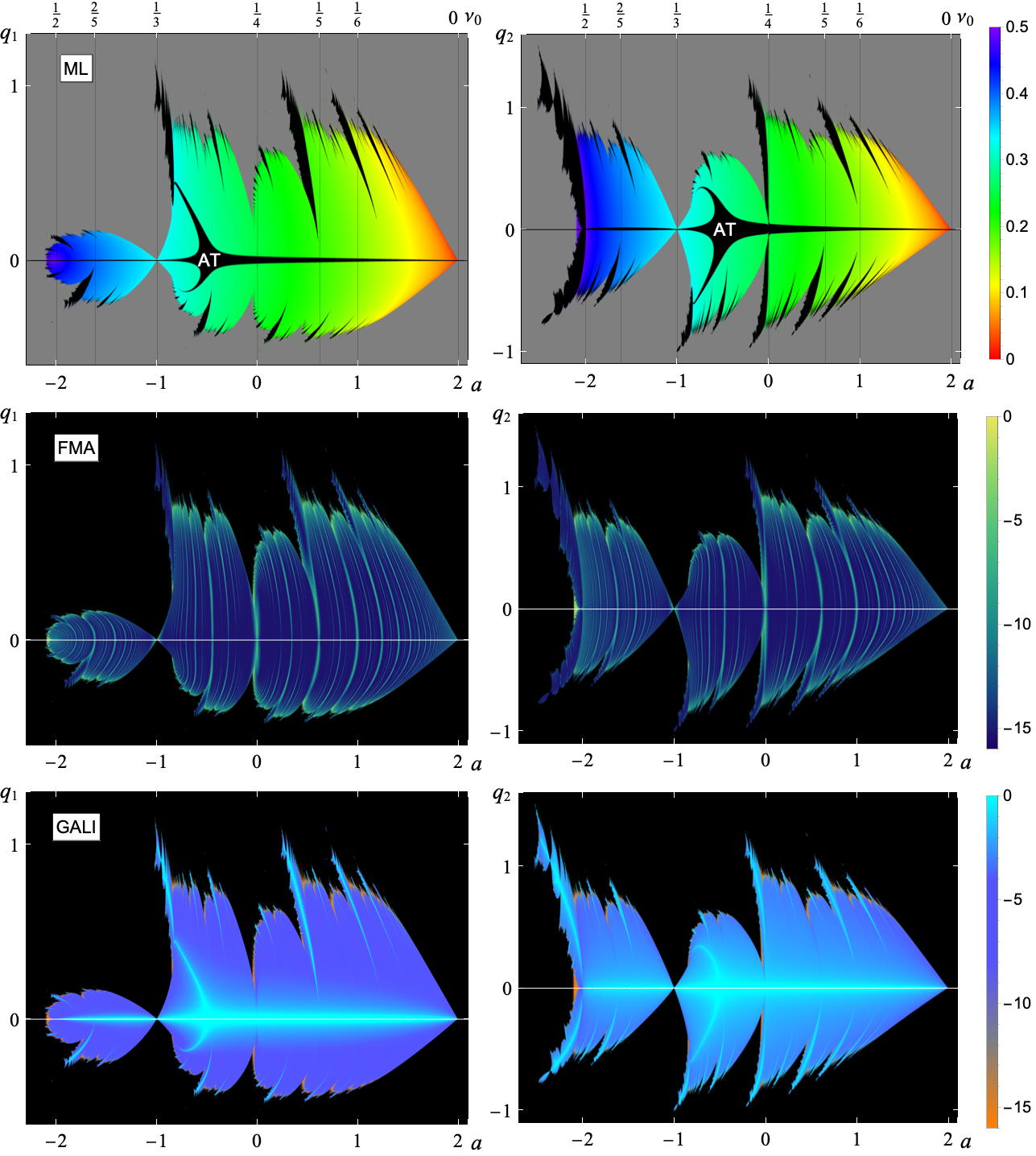}
    \caption{\label{fig:FractalNG}
    {\bf Isochronous and Period-doubling diagrams}.
    The left and right columns display stability diagrams along
    the first $l_1$ (isochronous) and second $l_2$ (period-doubling)
    symmetry lines, respectively.
    In each plot, the abscissa represents the transformation trace
    $a$ at the origin, while the ordinate corresponds to the
    horizontal coordinate along the symmetry line.
    Different rows present various indicators: rotation number with
    mode-locking (in black) (ML), frequency map analysis (FMA), and
    the generalized alignment index (GALI).
    Unstable orbits are shown in gray in the top plots and in black
    in the middle and bottom rows.
    }
\end{figure*}

Transitioning from phase-space plots to stability diagrams,
Figure~\ref{fig:FractalNG} demonstrates initial conditions
along both symmetry lines, colored by different dynamical
indicators.
The top row shows a mode-locking (ML) plot highlighting the
rotation number;
here, filtering emphasizes mode-locked regions in black,
with solid and dashed lines denoting isolated 1-, 2-, 3- and
4-cycles.
The rows below depict stable trajectories, with colors
corresponding to $\log_{10}(10^{-16} + ind)$ for both FMA and GALI
indicators.
For the FMA indicator, $ind_\text{FMA} = |\nu_1 - \nu_2|$, where
$\nu_{1,2}$ are the rotation numbers calculated over the first and
second halves of the iteration window.
For the GALI indicator, the value is derived from the product of
singular values of the normalized tangent matrix after $n$
iterations.
Chaotic trajectories are characterized by $color\approx 0$
for FMA and $color\approx 10^{-16}$ for GALI/REM.

Each indicator reveals unique details.
The ML plot highlights Arnold tongues --- chains of islands ---
but lacks internal tongue structure detail, which is more visible
in GALI and REM plots.
Among these, REM has proven particularly fast and efficient for
phase-space and stability diagrams, so it will serve as our
standard method for most of the remaining analysis.
Starting from a given initial condition, the REM dynamical indicator
is evaluated by iterating the map in the forward direction and then
in the backward direction,
$ind_\text{REM} = \sqrt{(q_{fin}-q_{ini})^2-(p_{fin}-p_{ini})^2}$.
Due  to finite numerical precision, the final state will generally
deviate from the initial condition.
This effect is notably amplified for chaotic initial conditions,
resulting in a substantially larger return error in such cases.

One notable feature in the ML plot is the large region surrounding
the origin, labeled ``AT'' for anti-tongue.
For GALI and REM plots, this feature shrinks closer to a line,
while the FMA plot omits it. The following table schematically
outlines differences among different fractal coloring methods:
\[
\begin{array}{l|ccc}
                        &\text{ ML }&\text{ FMA }&\text{GALI/REM}   \\\hline
\text{Tongues}          & +   & +   & +           \\
\text{Tongue structure} & -   & \pm & +           \\
\text{Anti-tongue}      & \pm & -   & +
\end{array}
\]

In the next section, we discuss these features in detail, while
here we briefly explain the choice of the term ``anti-tongue.''
First, by refining the mode-locking threshold, we observe that
Arnold tongues stabilize into finite areas where advanced
indicators reveal intricate internal structures, while the
anti-tongue, in contrast, narrows down to a thin line.
Furthermore, while the Arnold tongue's center aligns with the
isolated stable $n$-cycle, the anti-tongue converges to a
twistless torus --- a continuous invariant curve in phase space.
Visually, in the mode-locking plot, the anti-tongue contrasts
with the V-shaped tongues, forming an inverted $\Lambda$-shape
when it touches a fixed point.
Finally, consistent with its name, the anti-tongue can
``annihilate'' upon collision with a real tongue, disappearing
in a bifurcation --- mirroring the behavior of
particle-antiparticle pairs in physics.

\newpage
\section{\label{sec:Understanding}Understanding the diagram}

\subsection{Circle map and mode-locking}

Before diving into the detailed analysis of our stability diagrams,
it's useful to revisit a specific dynamical system from the family
of circle endomorphisms introduced by
V. Arnold~\cite{Arnold2009,BoylandCircle,KuznetsovCircle}.
This system is often employed to explore how intrinsic variables,
such as the rotation number, change with system parameters,
providing a foundational understanding that will support more
advanced generalizations.
Additionally, as will be seen in the final section, it has a
direct connection to a particular case study of the Chirikov map.

The {\it standard circle map} is a one-dimensional transformation
of a circle onto itself:
\[
\mathrm{T}_{\nu_0,\epsilon}:\,
\theta' = \phi(\theta)\mod 2\pi,
\qquad\qquad
\theta \in \mathbb{S}^1 = [0;2\pi),
\]
where the map is computed modulo $2\pi$, and the iterative
function or two parameters is defined as:
\begin{equation}
\label{math:CircleMap}    
    \phi(\theta) = \theta + \Omega + \epsilon\,\sin{\theta},
    \qquad\qquad
    \Omega = 2\,\pi\,\nu_0.
\end{equation}
Here, $\nu_0 \in [0;1)$ represents the {\it bare}/{\it natural}
rotation number of an oscillatory system, while $\epsilon > 0$
corresponds to the {\it coupling strength}, describing the level
of externally applied nonlinearity.

\begin{figure}[t!]
    \centering
    \includegraphics[width=0.9\linewidth]{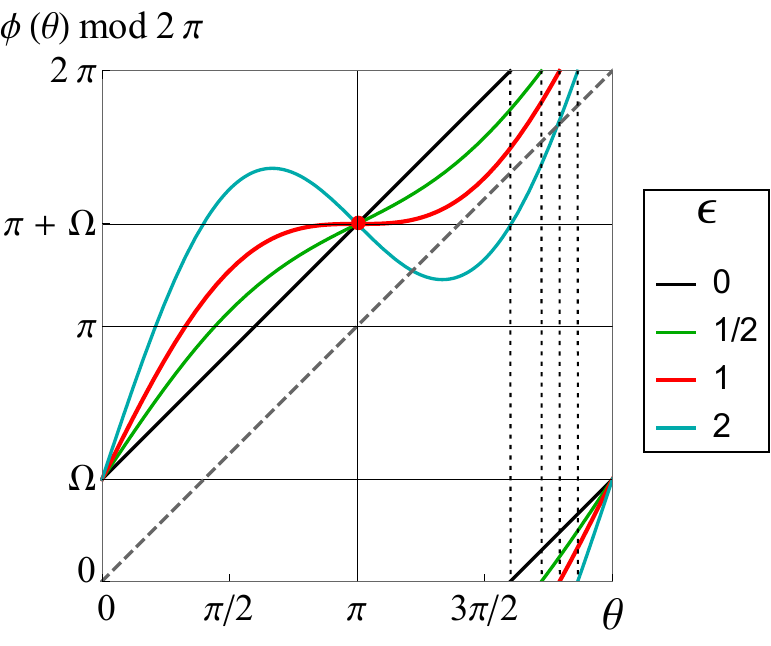}
    \caption{\label{fig:PhiTheta}
    {\bf Standard circle map's iterative function}.
    The solid colored curves represent one period of the 
    $\phi(\theta)\,\mathrm{mod}\,2\pi$
    for different values of the coupling strength parameter
    $\epsilon = 0,1/2,1,2$ as indicated in the legend.
    The red curve marks the critical case.
    The black dashed line corresponds to a linear function with
    a slope of 1, included for reference.
    }
\end{figure}

By analyzing the behavior of $\phi(\theta)$, we can identify key
scenarios depending on the value of $\epsilon$, with
Fig.~\ref{fig:PhiTheta} providing a visual representation:

\vspace{0.15cm}
\noindent $\bullet$ {\bf Unperturbed case}, $\epsilon = 0$.
The system exhibits a {\it rigid rotation}:
\[
    \theta' = \theta + \Omega,
\]
where every point moves at a constant angular velocity, $\Omega$.
This motion is non-chaotic and highly regular, with the rotation
number equal to $\nu_0$.
For rational values of $\nu_0 = p/q \in \mathbb{Q}$, the orbit
completes $p$ full rotations around the circle exactly after
$q$ iterations, forming a $q$-cycle.
For irrational values of $\nu_0$, the motion is quasiperiodic,
gradually covering the circle densely over time without ever
repeating in a strictly periodic pattern.

\vspace{0.15cm}
\noindent $\bullet$ {\bf Small perturbations}, $0 < \epsilon < 1$.
In this range, the function $\phi(\theta)$ remains monotonically
increasing (green curve in Fig.~\ref{fig:PhiTheta}),
meaning all orbits have to move forward.
The map $\mathrm{T}_{\nu_0,\epsilon}$ is an analytic
{\it diffeomorphism} --- smooth, invertible, and differentiable
(along with its inverse) transformation.
This case is particularly interesting to us, as it provides a
qualitative framework for interpreting our diagrams.

\vspace{0.15cm}
\noindent $\bullet$ {\bf High perturbations}, $\epsilon > 1$.
When $\epsilon$ exceeds 1, the function $\phi(\theta)$ is no longer
bijective (as shown by the cyan curve in Fig.~\ref{fig:PhiTheta}),
making the circle map $\T_{\nu_0,\epsilon}$ noninvertible.
This opens the possibility to more complex dynamics, such as
bistability and subharmonic routes to chaos.

\noindent $\bullet$ {\bf Critical case}, $\epsilon = 1$.
The value $\epsilon = 1$ is referred to as {\it critical}, as it
marks the boundary between two qualitatively different behaviors
seen in the intervals $0 < \epsilon < 1$ and $\epsilon > 1$.
The map becomes an analytic {\it homeomorphism} of the circle,
featuring a single cubic {\it critical point} where
$\phi(\theta)$ is still continuous and invertible, but its
inverse is no longer smooth (refer to red point/curve
in Fig.~\ref{fig:PhiTheta}).

\begin{figure}[t!]
    \centering
    \includegraphics[width=0.85\linewidth]{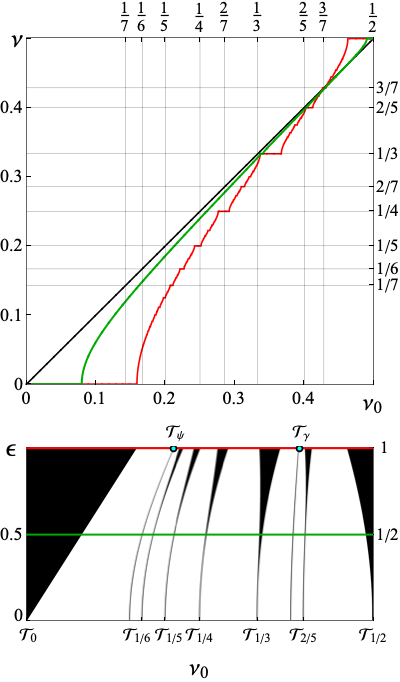}
    \vspace{-0.2cm}
    \caption{\label{fig:devils}
    {\bf Arnold tongues}.
    The top plot shows examples of devil's staircases, where the
    rotation number $\nu$ is plotted against the bare tune $\nu_0$,
    with constant values of $\epsilon=1$ (red) and $\epsilon=1/2$
    (green).
    The bottom plot schematically illustrates several Arnold tongues
    in the $(\nu_0, \epsilon)$ space, shown in black.
    Rational tongues $\mathcal{T}_\alpha$ corresponding to
    mode-locking are labeled along the horizontal axis, where $\alpha$
    matches the value of $\nu_0$.
    Two curves associated with irrational tongues, for
    $\psi=\sqrt[5]{2}-1$ and $\gamma = 2-\phi$ (with $\phi$ as the
    golden ratio), are marked at the top, with terminal points
    indicated in cyan.
    The red and green lines link the devil's staircases in the top
    plot to their corresponding parameter spaces below.
    }\vspace{-0.5cm}
\end{figure}

In the space of the map's parameters, for each orbit with initial
condition $\theta_0$, the rotation number (if it exists) is given
by the limit:
\[
\nu = \frac{1}{2\,\pi}\,\lim_{n\rightarrow\infty}
    \frac{\phi^n(\theta_0)-\theta_0}{n}.
\]
The set of parameters corresponding to a given rotation number
$\alpha$,
\[
\mathcal{T}_\alpha = \{(\nu_0,\epsilon)|\,\nu = \alpha\}
\]
is known as the {\it Arnold $\alpha$-tongue}.
For any fixed value of $0 < \epsilon \leq 1$, the rotation number 
$\nu$ as a function of $\nu_0$ forms a continuous, non-decreasing
curve known as the ``devil’s staircase,'' with flat steps of
non-zero width at each rational value of $\nu$.
These flat regions correspond to {\it mode-locked} states, where
the dynamics are periodic, and the system is said to be ``locked''
to a specific rational rotation number.
The boundaries of these regions are determined by algebraic curves
satisfying the periodic condition for orbits with unit multipliers.
Two examples of this staircase for $\epsilon = 1/2$ (green) and 
$\epsilon = 1$ (red) are shown in the top plot of
Fig.~\ref{fig:devils}.
For $\epsilon = 0$, the staircase degenerates into a linear function
(black), with the flat segments disappearing entirely, as the set
of rational numbers in the interval $[0,1]$ has a total measure of
zero.

As $\epsilon$ increases, the measure of mode-locked states grows
from zero to one, leading to a transition where quasiperiodic
states become rare at $\epsilon=1$.
At this critical point, the staircase becomes ``complete,'' meaning
it is constant almost everywhere except on a Cantor set, with a
fractal dimension of approximately $\approx 0.870$.
This transition creates the familiar V-shaped regions, or rational
tongues, in the parameter space, as shown in the bottom plot of
Fig.~\ref{fig:devils}.
For a fixed value of $\epsilon$, rational tongues can be ordered
by width, following a Farey sequence.

\begin{figure}[b!]
    \centering
    \includegraphics[width=\linewidth]{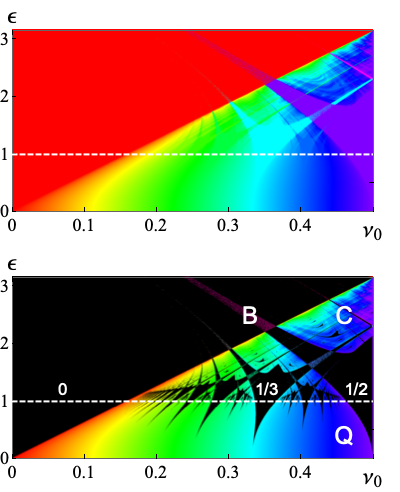}
    \caption{\label{fig:ATongues}
    {\bf Dynamics of Circle map}.
    The top plot shows the rotation number for the standard circle
    map across its parameter space $(\nu_0, \epsilon)$, with $\nu$
    evaluated for randomly chosen initial conditions $\theta_0$.
    The bottom plot classifies the dynamical regimes, marked with
    white labels.
    Black regions represent mode-locked states (some tongues are
    labeled with their corresponding rotation numbers), with ``noisy''
    strips of constant color within these areas indicating regions of
    bistability (B).
    Colored points represent quasiperiodic motion (Q) for $\epsilon<1$
    and chaotic motion (C) for $\epsilon > 1$.
    The horizontal axis presents the color scale, since $\nu=\nu_0$
    when $\epsilon = 0$.
    }
\end{figure}

While rational tongues have an interior, tongues with irrational
rotation numbers correspond to Lipschitz continuous curves
connecting points $(\alpha,0)$ with $(\alpha_t,1)$, where
$\alpha_t$ is a terminal value.
In the bottom plot of Fig.~\ref{fig:ATongues}, two constant level
sets of $\nu$ are shown for $\mathcal{T}_{\psi,\gamma}$, where
$\psi=\sqrt[5]{2} - 1$ and $\gamma=2-\varphi=(3-\sqrt{5})/2$
(with $\varphi$ being the golden ratio and the golden mean (GM)
critical point at $\alpha_t \approx 0.607$).

For $\epsilon > 1$, rational tongues may overlap, leading to
complex behaviors such as bistability, where the system can exhibit
two distinct stable states depending on the initial conditions, or
period-doubling routes to chaos.
Fig.~\ref{fig:ATongues} provides further details on these different
dynamical regimes.

The top plot in Fig.~\ref{fig:ATongues} shows the parameter space,
where the color represents the rotation number $\nu$ for initial
conditions randomly selected from $\theta_0\in[0,2\pi)$.
In contrast, the bottom plot uses additional black color to indicate
mode-locked regions, offering a clear visualization of different
dynamical regimes.
The lower plot, with $\epsilon$ below the critical threshold, now
offers a qualitative ``zeroth-order'' description of our diagrams
in Fig.~\ref{fig:FractalNG} for amplitudes below the resonance
overlap, corresponding to the loss of stability.
We will now explore the differences and proceed with a more detailed
analysis.
\vspace{-1cm}

\newpage
\begin{figure}[t!]
    \centering
    \includegraphics[width=\linewidth]{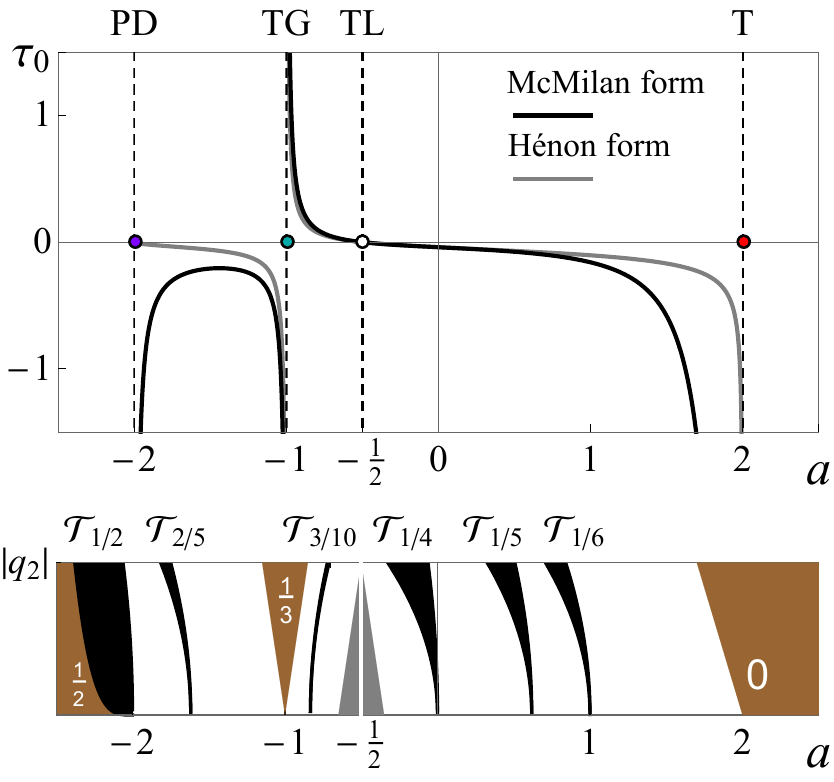}
    \caption{\label{fig:dNUdJ}
    {\bf H\'enon twists}.
    The top plot shows the twist coefficient $\tau_0(a)$ for
    the H\'enon quadratic map in both McMillan (black) and H\'enon
    (gray) forms.
    The bottom plot schematically illustrates the structure of
    Arnold tongues along the second symmetry line, $l_2$, for small
    amplitudes ($q_2 < 0$).
    Regular tongues are shown in black, with the anti-tongue in
    gray.
    For remaining initial conditions, stable regions (both rational
    and irrational) are shown in white, while unstable regions
    (singular tongues) are
    marked in gold, with numbers indicating the resonance values of
    $\nu_0$ associated with instability.
    }
\end{figure}

\subsection{\label{sec:twisted}Twisted tongues}

When comparing the ML diagrams in Fig.~\ref{fig:FractalNG} with
the bifurcation diagram of the circle map Fig.~\ref{fig:ATongues},
two qualitative differences stand out:
\begin{itemize}
    \item[(i)] Although rational tongues again appear in a sequence
    similar (but different) to the Farey sequence, for fixed values
    of $\nu_0 = 0$ and $\nu_0 = 1/3$ with small but nonzero
    distances along the symmetry lines ($|q_{1,2}| > 0$), we observe
    instability rather than mode-locked motion --- {\it singular
    tongues}.
    For $\nu_0 = 1/2$ (with $a=-2$), a pair of tongues
    $\mathcal{T}_{1/2}$ forms along the second symmetry line;
    however, the motion becomes unstable near the origin.
    \item[(ii)] Tongues respond differently to perturbations by
    varying slopes as a function of amplitude.
    For small $\epsilon$ in the circle map, the rotation number's
    derivative with respect to $\epsilon^2$ varies monotonically
    \[
        \lim_{\epsilon\rightarrow 0}\dd(2\,\pi\,\nu)/\dd(\epsilon^2) = -\frac{1}{4}\,\cot[\pi\,\nu_0].
    \]   
    In contrast, for the quadratic H\'enon map, with small
    $|q_{1,2}|$, tongues within $\nu_0\in(\nu_\chi,1/3)$
    ($-1 < a < -1/2$) lean toward $\nu_0 = 0$, while otherwise
    they lean toward $\nu_0 = 1/2$.
\end{itemize}
Figure~\ref{fig:dNUdJ} very schematically depicts the Arnold tongue
structure, with the vertical axis representing $|q_2|$ for $q_2<0$.
Stable (rational and irrational) trajectories appear in white,
while black regions indicate some mode-locked areas, and gold
shows unstable initial conditions.

These differences can be understood through foundational works in
dynamical systems and chaos theory, especially in the context of
the H\'enon map.
We refer readers to relevant articles
~\cite{dulin2000henon,dullin2000twistless,sterling1999homoclinic}
and their references for further detail, while we briefly
summarize some key results here.
According to the KAM theorem
\cite{kolmogorov1954conservation,moser1962invariant,arnol1963small},
most invariant tori near a stable fixed point persist under small
perturbations.
This allows us to define the map’s {\it canonical form} as
\[
\begin{array}{l}
    J' = J,                 \\[0.25cm]
    \theta' = \theta + 2\,\pi\,\nu(J),
\end{array}
\]
where $J$ is the {\it action} (or {\it symplectic radius}) and
$\theta$ is the {\it conjugate angle} variables.
In Birkhoff normal form theory (see, e.g.,\cite{dullin2000twistless}),
the rotation number is often expressed as a power series of the
action:
\[
\nu(J) = \nu_0 + \tau\,J
    = \nu_0 + \tau_0 J + \frac{1}{2!}\,\tau_1 J^2
        + \frac{1}{3!}\,\tau_2 J^3 + \mathcal{O}(J^4),
\]
where the derivative of $\nu$ with respect to $J$
\[
\tau(J) = \frac{\dd \nu}{\dd J} = \tau_0 + \tau_1 J +
        \frac{1}{2}\,\tau_2 J^2
        + \mathcal{O}(J^3)
\]
known as the {\it twist}, plays a critical role in nonlinear
stability.

For the McMillan form mappings with a smooth, differentiable force
function expandable around the origin as:
\[
    f(q) = a\,q + b\,q^2 + c\,q^3 + \ldots
\]
the first twist coefficient, $\tau_0$, is expressed as:
\begin{equation}
\label{math:twist}
2\,\pi\,\tau_0 = \frac{1}{4-a^2}\,\left[
    4\,b^2\frac{a+1/2}{(a-2)(a+1)} - 3\,c
\right].
\end{equation}
When $b \neq 0$, $\tau_0$ is defined for values of $\nu_0$
excluding $0,1/2$ and $1/3$ ($a \neq 2,-2,-1$), where it becomes
singular, while $\tau_1$ also requires $\nu_0\neq 1/4,1/5,2/5$
($a\neq 0,(-1\pm\sqrt{5})/2$).
This result can be independently derived through various methods,
such as Birkhoff normal forms~\cite{dullin2000twistless}, Lie
algebra~\cite{bengtsson1997, morozov2017dynamical}, square matrix method~\cite{hua2017square}, Deprit
perturbation theory~\cite{michelotti1995intermediate}, and, as
recently shown, also extracted from related integrable McMillan
multipoles~\cite{zolkin2024MCdynamics,zolkin2024MCdynamicsIII},
which offer additional qualitative insights.
The top plot in Fig.~\ref{fig:dNUdJ} shows $\tau_0(a)$ for the
quadratic H\'enon map in both forms of the transformation.

When the rotation number of the stable fixed point reaches a
rational value $m/n$ and the isoenergetic nondegeneracy condition
$\tau_0 \neq 0$ is met, KAM theory provides the following
qualitative picture:
\begin{itemize}
    \item For $n=1,2$ the fixed point loses linear stability when
    $|a|$ crosses 2.
    At $\nu_0 = 0$, the quadratic H\'enon map undergoes a
    transcritical (T) bifurcation, while a cubic map undergoes a
    symmetric pitchfork (PF), creating an additional pair of fixed
    points.
    In Fig.~\ref{fig:dNUdJ}, we see that $\tau_0$ is singular or
    zero at the period-doubling bifurcation (PD), depending on the
    from of the map.
    This difference explains why the fractal structure in the original
    H\'enon map extends infinitely near $\nu_0=1/2$ while in the
    McMillan form it remains finite.
    \item For $n=3$, generally, the resonant term dominates the
    twist term near the origin, causing instability.
    In the quadratic H\'enon map, this results in a touch-and-go
    (TG) bifurcation, while cubic or Chirikov mappings, due to
    additional spatial symmetry, stabilize with a symmetric pair
    of period-3 chains.
    \item For $n=4$ the origin may become unstable if the resonant
    term is sufficiently large.
    \item For $n>4$, the Moser Twist theorem implies stability of
    the fixed point, with bifurcations resulting in the creation
    of pairs of elliptic and hyperbolic $n$-cycles.
    Additionally, if $\tau_0=0$ and low-order resonances ($n\leq 6$)
    are absent, $\tau_1 \neq 0$ provides a necessary stability condition~\cite{siegelmoser1971,arnold1988,arnold2006}.
\end{itemize}

When the twist vanishes at the fixed point, the stability
condition becomes more complex, leading to what is referred to as
a {\it twistless} (TL) bifurcations.
As noted in~\cite{dulin2000henon}, for the generalized H\'enon map,
assuming $n > 6$ and $b \neq 0$, $\tau_1$ remains non-zero when
$\tau_0 = 0$.
From Eq.~(\ref{math:twist}), we can understand the specific
evolution of tongues around the origin, as well as why for the
quadratic H\'enon map we expect a twistless bifurcation at $a=-1/2$,
which corresponds to the middle of the anti-tongue with an
irrational rotation number:
\[
    \nu_\chi = \frac{1}{2\,\pi}\,\arccos[-1/4].
\]
In contrast, the cubic map does not exhibit a TL bifurcation at $q=0$,
because $\tau_0$ maintains a constant sign and never vanishes.

\begin{figure}[h!]
    \centering
    \includegraphics[width=\linewidth]{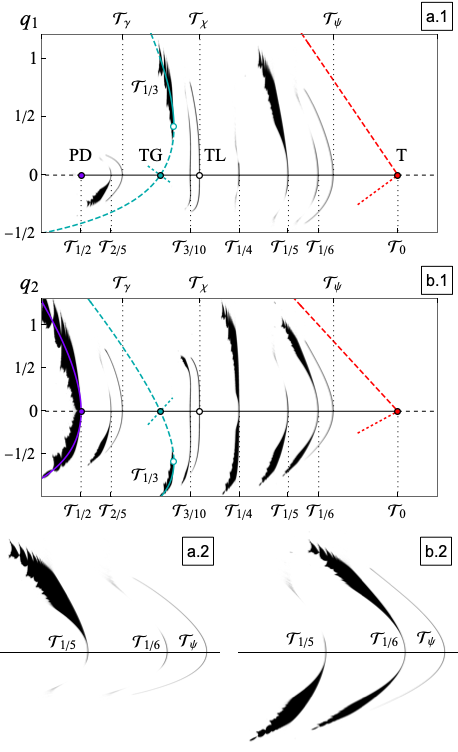}\vspace{-0.25cm}
    \caption{\label{fig:HTongues}
    {\bf Feathers}.
    Structure of Arnold tongues along the first (a.1) and second
    (b.1) symmetry lines for the quadratic H\'enon map: rational
    tongues at
    $\nu_0 = \frac{1}{2},\frac{1}{3},\frac{1}{4},\frac{1}{5},\frac{2}{5},\frac{1}{6}$,
    and $\frac{3}{10}$, and irrational tongues at
    $\nu_0 = \psi,\gamma$, and $\chi$ (associated with the TL
    bifurcation).
    Singular and twistless bifurcations of the stable fixed point
    are indicated by points along the horizontal axis:
    transcritical (T), period-doubling (PD), touch-and-go (TG),
    and the creation of twistless torus (TL).
    Dashed lines depict the coordinates for the unstable
    fixed point (red), 2-cycle (purple), and 3-cycles
    (cyan, solid when stable).
    Dotted lines of the same colors indicate approximate locations
    of crossings with corresponding stable/unstable manifolds.
    The bottom plots magnify the vicinity of resonances at
    $\nu_0 = 1/5$ and $\nu_0 = 1/6$, highlighting the typical
    structures for odd and even island chains.
    }\vspace{-0.25cm}
\end{figure}

Figure~\ref{fig:HTongues} displays several rational and irrational
tongues obtained by extracting constant level sets from the rotation
number plots.
Colored points and lines, mark singular bifurcations at the origin
and represent unstable $n$-cycles, while the white point indicates
the twistless bifurcation.
The magnified plots (a and b.2) illustrate tongues for even
($\nu_0=1/6$) and odd ($\nu_0=1/5$) chains, revealing self-similar
fractal structures resembling feathers, typical for higher-order
resonances outside the positive twist region $a\in(-1,-1/2)$
(i.e., $\nu_0\in(\nu_\chi,1/3)$).

\begin{figure*}[t!]
    \centering
    \includegraphics[width=\linewidth]{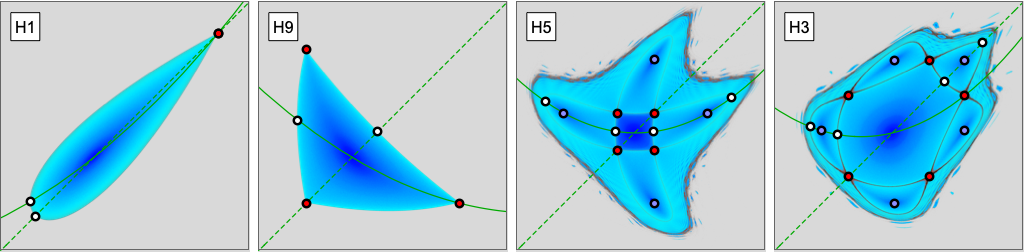}
    \caption{\label{fig:EvolutionEX1}
    {\bf Symmetry lines and tongues}.
    The first two plots show phase-space portraits near singular
    tongues corresponding to transcritical [H1] and touch-and-go [H9]
    bifurcations, where $\nu_0$ is close to 0 and 1/3.
    The following two plots represent typical scenarios for regular
    tongues, illustrating even [H5] and odd [H3] island bifurcations
    near $\nu_0 = 1/4$ and $\nu_0 = 1/5$.
    Stable cycles are shown in blue, unstable cycles in red, and
    intersections of symmetry lines with stable/unstable manifolds
    are highlighted in white.
    These phase-space portraits are recreations of H\'enon's original
    palettes (Fig.~\ref{fig:FractalOG}), presented here in McMillan
    form, with REM parameters used for coloring.
    }\vspace{-0.4cm}
\end{figure*}

Notably, unlike the circle map, where every tongue near
$|\epsilon| \approx 0$ is V-shaped, the situation differs here.
As previously discussed, on one side of each symmetry line
(for odd chains) or at both ends of one of the symmetry lines
(for even chains), we encounter unstable nodes of the chain
instead of crossing mode-locked islands.
Consequently, for odd chains, the ``feathers'' are paired with
lines, referred to as ``cuts.''
For even chains, two feathers or two lines appear together. 
In the next subsection, we will further investigate these
crossings, while Figure~\ref{fig:EvolutionEX1} provides several phase
space plots that visually support the diagrams in
Figure~\ref{fig:HTongues}.

\subsection{Tears and frays}

\begin{figure}[b!]
    \centering
    \includegraphics[width=0.9\linewidth]{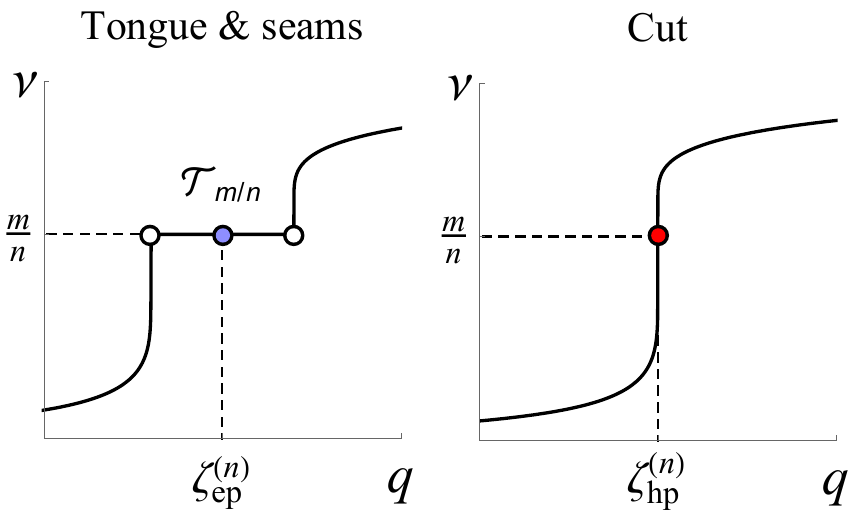}\vspace{-0.4cm}
    \caption{\label{fig:TSC}
    {\bf Seam and cut}.
    Schematic illustration shows the devil's staircase pattern as
    the symmetry line crosses through the center of the island (stable
    $n$-cycle $\z^{(n)}_\text{ep}$ in blue) and node (unstable
    $n$-cycle $\z^{(n)}_\text{un}$ in red).
    The left plot represents an Arnold tongue --- a flat region
    bordered by intersections with the stable and unstable manifolds
    of $\z^{(n)}_\text{un}$ (white points).
    In the right plot, a ``cut'' appears where the derivative of
    $\nu(q)$ diverges to infinity.
    }
\end{figure}

Analyzing the rotation number along the symmetry lines, we observe
a flat region when crossing an island center, similar to the
behavior seen in the circle map (left plot in Fig.~\ref{fig:TSC}).
However, at the node crossing, no mode-locking region appears, and
the derivative of $\nu$ diverges to infinity (right plot).

\begin{figure*}[h!]
    \centering
    \includegraphics[width=\linewidth]{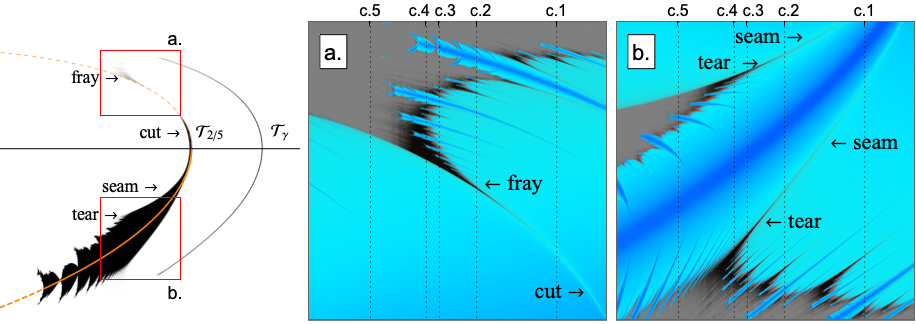}\vspace{0.05cm}
    \includegraphics[width=\linewidth]{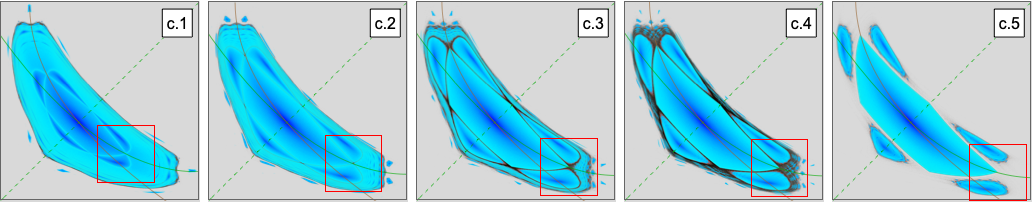}\vspace{0.05cm}
    \includegraphics[width=\linewidth]{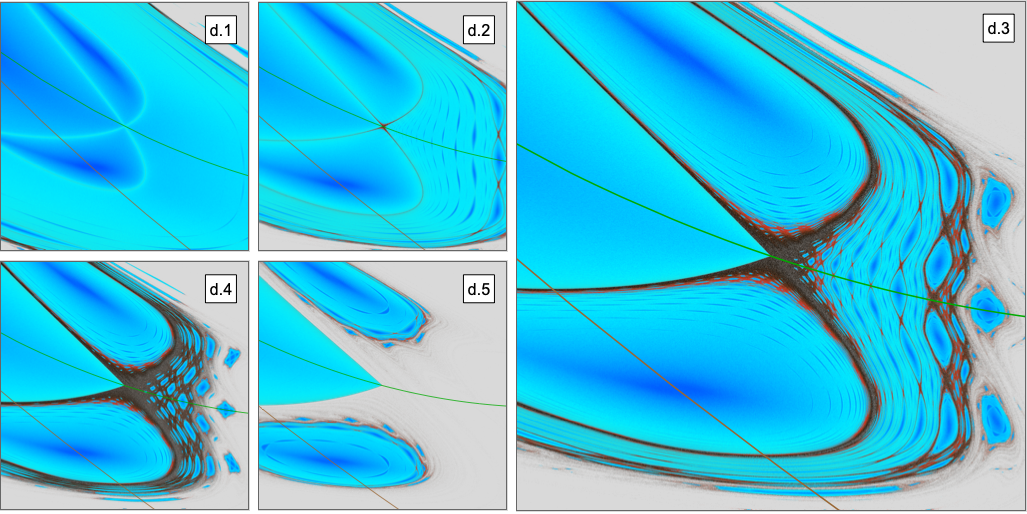}
    \caption{\label{fig:Fray}
    {\bf Tear and fray}.
    Schematic illustration of the rational $\mathcal{T}_{2/5}$ and
    irrational $\mathcal{T}_\gamma$ Arnold tongues along the first
    symmetry line of the quadratic H\'enon map.
    Points in black mark the level sets for these rotation numbers,
    determined with accuracies of $\delta\nu_{2/5} = 3 \times 10^{-4}$
    and $\delta\nu_\gamma = 8 \times 10^{-5}$.
    Orange lines represent branches of the 5-cycle, shown as solid
    for stable and dashed for unstable.
    Insets (a.) and (b.) display REM indicators within the areas
    enclosed by red rectangles, highlighting features: seams and cuts
    in light blue, frays and tears in red/dark, rational and irrational
    rotations in shades of blue, and unstable regions in gray.
    Palettes (c.1--5) show phase-space diagrams and magnifications
    (d.1-5), illustrating typical stages in the destruction of invariant
    tori.
    Alongside the primary symmetry lines $l_{1,2}$ (green), a brown line
    represents the inverse symmetry $l_2^{-1}$, mirroring the crossing
    behavior on the opposite side of $l_2$.
    }
\end{figure*}

As the system shifts toward chaotic behavior with overlapping
resonances, the appearance of both tongues and cuts changes
distinctly.
To illustrate, we use an odd chain of 5 islands, chosen because
it intersects both stable and unstable cycles on opposite sides
of each symmetry line, making it convenient for study.
The left plot in Fig.~\ref{fig:Fray} shows two constant level
sets, one for the rational tongue $\mathcal{T}_{2/5}$ and
one for the irrational tongue $\mathcal{T}_\gamma$ along $l_1$,
extracted with accuracies of $\delta\nu_{2/5} = 3\times10^{-4}$
and $\delta\nu_\gamma = 8\times10^{-5}$, respectively.

In the upper part of the plane (for $q_1>0$), the constant level
set $\nu = 2/5$ splits into two distinct sections:
a ``cut,'' tracing an unstable 5-cycle (dashed orange line), which
transitions into a ``fray.''
Both terms draw an analogy to tailoring, with the cut splitting
the body of the main fractal.
As the system transitions to chaos, this once fine line now
resembles fraying fabric, with threads coming apart to form
a loose edge.
With increasing numerical precision, the cut narrows, exhibiting
behavior characteristic of an irrational tongue, while the fray
maintains its structure, showing chaotic orbits slowly wandering
around the high-order resonances.

On the opposite side of the symmetry line, where $l_1$ crosses
the middle of the island, the tongue takes on a wedge-like shape,
similar to a dart often seen in sewing patterns.
The middle of the dart represents the stable $5$-cycle (solid
orange line), with edges attached to the main fractal by ``seams,''
much like two pieces of fabric sewn together.
These seams mark the intersections of stable and unstable manifolds,
forming an isolating separatrix that confines motion below the
resonance overlap.
When motion becomes chaotic, the seams detach from the main fractal, causing a ``tear.''

The two plots on the right display the REM indicator for sections
of the H\'enon set near the tongue structure.
Plot (a.) magnifies the fray, while plot (b.) reveals the seams
and tears in the structure.
In the chosen color scheme, gray represents unstable trajectories,
red and dark shades indicate chaotic orbits, light blue highlights
the cut and seam, and various shades of blue represent irrational
KAM circles, with the color darkening around stable $n$-cycles.
Palettes (c) show phase-space portraits for typical resonance
stages: plot (c.1) and (c.2) showcase situation before and near
resonance overlap, plot (c.3) illustrates strong overlap with
some isolating invariant structures remaining, and plots (c.4,5)
depicts the scenario where no invariant tori prevent escape.
Finally, palettes (d.1–5) provide detailed magnifications of the
island structure along the second symmetry line and its inverse,
which mimics the opposite side of the tongue.

\begin{figure*}[t!]
    \centering
    \includegraphics[width=\linewidth]{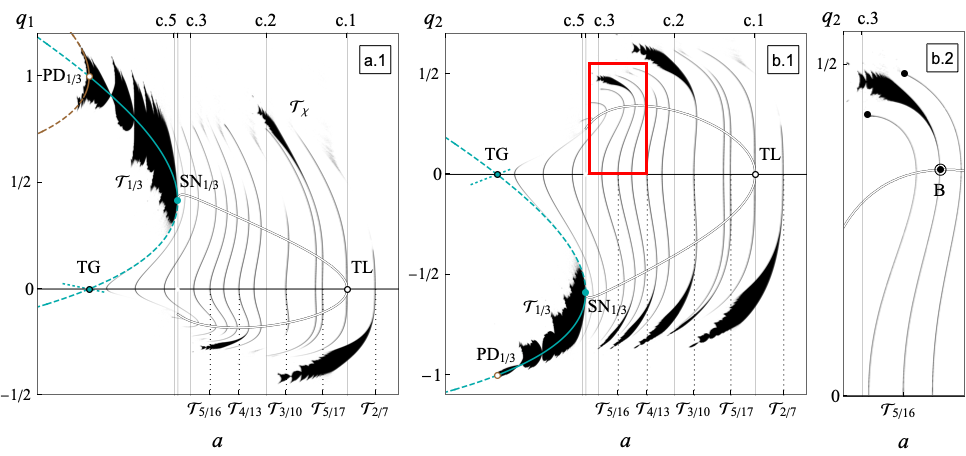}\vspace{-0.4cm}
    \caption{\label{fig:Cobra}
    {\bf Ribcage and cobras}.
    Structure of Arnold tongues in the H\'enon map for the region
    with positive twist coefficient $\tau_0 > 0$, i.e.,
    $a \in (-1, -1/2)$.
    Plots (a.1) and (b.1) highlight rational tongues along the first
    and second symmetry lines at
    $\nu_0=\frac{2}{7},\frac{5}{17},\frac{3}{10},\frac{4}{13},\frac{5}{16}$,
    and $\frac{1}{3}$; all other tongues, including $\mathcal{T}\chi$,
    represent irrational values of $\nu_0$.
    Colored points mark singular (TG) and twistless (TL)
    bifurcations at the origin and a saddle-node ($\text{SN}_{1/3}$)
    bifurcation where the twistless orbit (black/white curve)
    collides with stable and unstable 3-cycles (solid and dashed
    cyan curves).
    At $a = -1$, the stable 3-cycle undergoes a period-doubling
    bifurcation, producing a pair of stable 6-cycles (brown curve).
    The right plot (b.2) magnifies the red rectangle region in
    (b.1), detailing the rational tongue $\mathcal{T}_{5/16}$ and
    two irrational tongues (curves ending at terminal points).
    A bifurcation (B) occurs when a rational tongue intersects
    the twistless orbit.
    Markers at the top (c.1--5) indicate parameter values for
    corresponding phase-space diagrams and rotation number
    dependencies on $q_{1,2}$ (d.1--5), as provided in
    Fig.~\ref{fig:FreqAT}.
    }
\end{figure*}

\newpage

\subsection{\label{sec:twistless}Twistless torus}

We now shift our attention to the region with a positive twist
coefficient, as illustrated in Fig.~\ref{fig:Cobra}, which
highlights several rational and irrational level sets of $\nu_0$.
In this region, the rotation number becomes a non-monotonic
function of amplitude, exhibiting a local minimum at the origin
and maxima at the locations of the twistless orbit.
This behavior causes the tongues to change the slopes of their
boundaries, giving them a ``cobra-like'' appearance (see plot
b.2) rather than the feather-like structure observed in
Fig.~\ref{fig:HTongues}.
When a tongue crosses a twisted torus (marked by point B),
higher-order bifurcations, such as saddle-node or reconnection
bifurcations, are expected
(for a detailed discussion, see~\cite{dullin2000twistless}).
Observing plots (a.1) and (b.1), the system resembles an X-ray of
a fish: the twistless torus (depicted as the black/white curve)
forms a ``ribcage,'' the fixed point at the origin acts as a
backbone, and the surrounding tongues represent ribs.
While twistless tori are generally associated with bifurcations,
it appears to have a stabilizing effect in the H\'enon quadratic
map for moderate amplitudes.
Within the region bounded by the twistless torus, tongues are
significantly narrower (potentially lines), resulting in
quasi-integrable dynamics.
However, this stabilizing effect can be disrupted if the twistless
orbit intersects major resonances, as will be discussed in the
final section.

\begin{figure*}[t!]
    \centering
    \includegraphics[width=\linewidth]{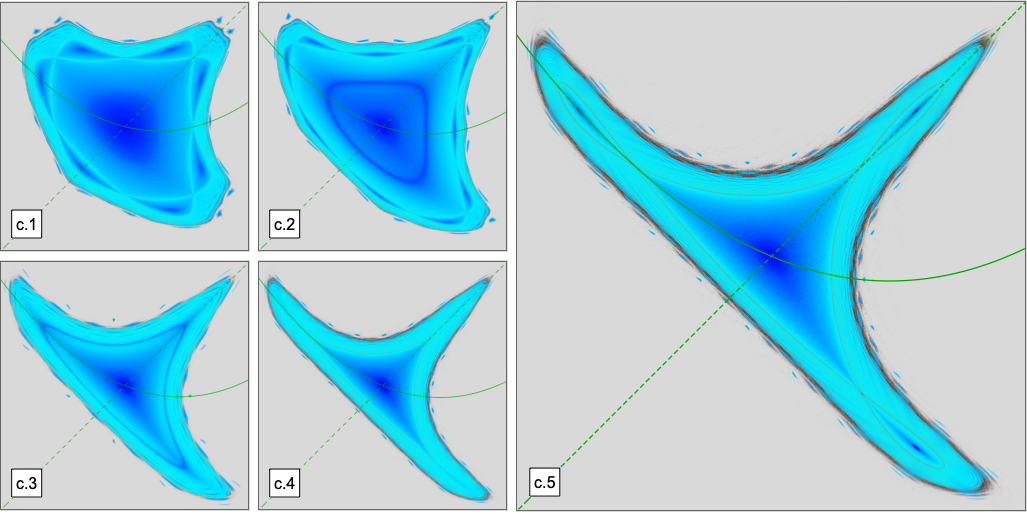}
    \includegraphics[width=\linewidth]{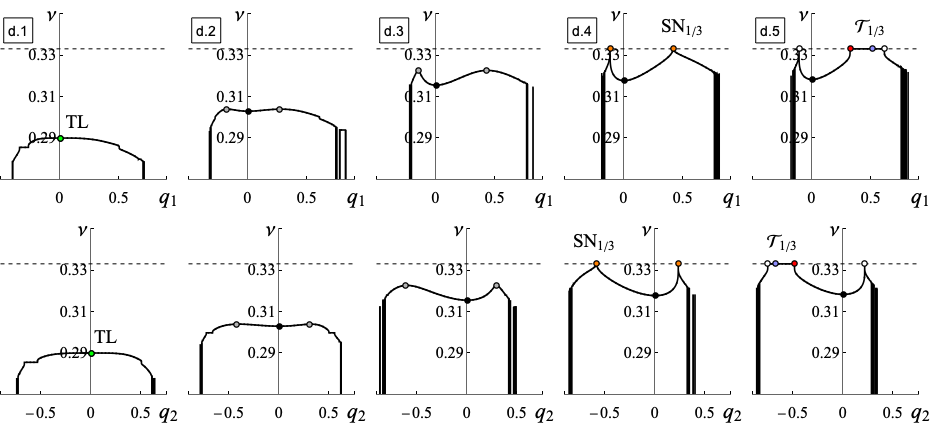}
    \caption{\label{fig:FreqAT}
    {\bf Evolution of the twistless orbit}.
    The top group of plots shows phase-space portraits at different
    stages of the twistless orbit's evolution
    (see Fig.~\ref{fig:Cobra} for further details):
    its birth at the origin in a twistless (TL) bifurcation (c.1),
    separation from the origin (c.2),
    just before cusp formation (c.3),
    annihilation in a saddle-node ($\text{SN}_{1/3}$) with a pair
    of 3-cycles (c.4),
    and the subsequent formation of 3 islands (c.5).
    Before bifurcation, the orbit is shown in dark blue,
    shifting to light blue at $\nu_0=1/3$.
    The bottom rows of plots (d.1--5) show rotation number
    dependencies along both symmetry lines (solid and dashed
    green curves).
    Colored points mark the twistless orbit (gray), the cusp orbit
    (orange), and stable and unstable 3-cycles (blue and red,
    respectively).
    }
\end{figure*}

Tick marks at the top (c.1--5) correspond to phase-space portraits
presented in Fig.~\ref{fig:FreqAT}, illustrating the evolution of
the twistless orbit.
Here, again, the REM indicator is used as a color scale: dark blue
represents orbits near stable $n$-cycles and the twistless orbit
itself (visible as a dark blue ring in palettes c.2 and c.3), light
blue indicates less regular orbits, including unstable cycles and
their manifolds, while red and dark colors represent chaotic
trajectories.

At $a = -1/2$ ($\nu_0 = \chi \approx 0.290$), a stable (since
$\tau_1 \neq 0$) twistless bifurcation occurs, see palette (c.1).
As $a$ decreases, a twistless torus detaches from the origin (c.2),
corresponding to an orbit with a local maximum in the rotation
number.
The bottom rows (d.1--5) complement the phase-space diagrams by
providing rotation number plots $\nu(q_{1,2})$.
Discontinuities and noise at larger amplitudes indicate instability
and chaos.

As $a$ approaches $2 - 2\sqrt{2}$, the twistless orbit deforms
(c.3) and develops a cusp at $\nu_0 \approx 0.318$ (c.4).
The cusp consists of a single parabolic 3-cycle and its associated
manifolds.
This configuration eventually breaks into a pair of stable and
unstable 3-cycles (c.5) in a saddle-node bifurcation
($\mathrm{SN}_{1/3}$).
At $a = -1$ ($\nu_0 = 1/3$), the stable cycle loses stability via
a period-doubling ($\mathrm{PD}_{1/3}$) bifurcation, while the
unstable cycle undergoes a touch-and-go (TG) bifurcation, passing
through the origin and inducing instability at the resonance.

Examining the rotation number plots, we observe that as the system
approaches the saddle-node bifurcation (d.3), the twistless orbit
approaches $\nu_0 = 1/3$, creating a cut and seam around the origin,
marked with orange points (d.4).
After the saddle-node bifurcation ($a < -1$), one of the orange
points (corresponding to the crossing of heteroclinic manifolds)
remains as a cut in the vicinity of this parameter before
transitioning into a fray.
The other orange point evolves into a plateau corresponding to the
regular tongue $\mathcal{T}_{1/3}$ (d.5);
the blue and red points mark locations of stable and unstable
3-cycles respectively.

\begin{figure}[t!]
    \centering
    \includegraphics[width=\linewidth]{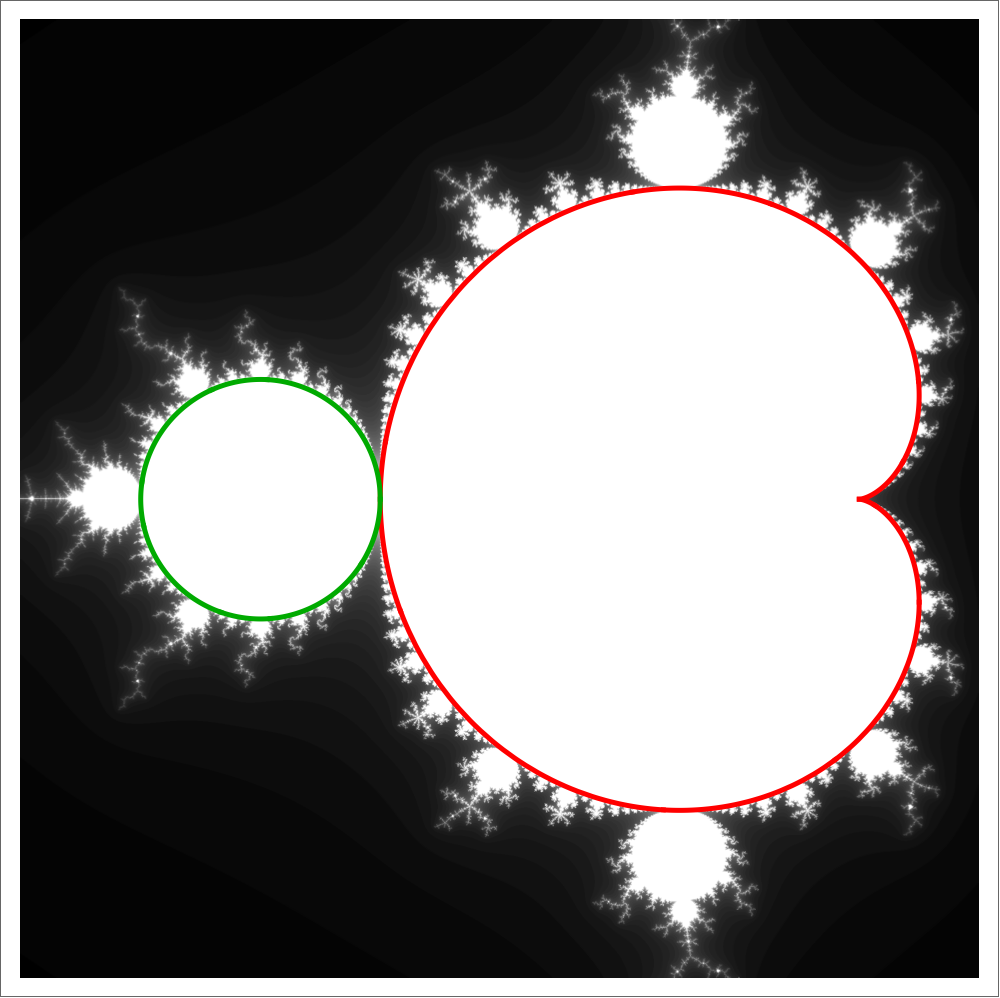}
    \caption{\label{fig:Cardioid}
    {\bf Elementary domains of the MBM set.}
    The Mandelbrot-Brooks-Matelski set with its central cluster
    approximated by a cardioid (red) and a cluster at a non-root
    node approximated by a circle (green).
    Adapted from~\cite{dolotin2008}.
    }
\end{figure}

\subsection{Shapes of Elementary Domains}

We began this section with a model problem: Arnold’s circle map.
This 1D system provided valuable qualitative insights into the
tongue structures near the origin.
When combined with our understanding of twist behavior in
symplectic maps, it also enabled reasonable quantitative
estimates.
However, as the system's complexity increases, the circle map
becomes insufficient for understanding the overall structure of
the H\'enon set, where critical amplitudes form complex
geometries unlike the simple horizontal critical line at
$\epsilon = 1$.

In contrast, the MBM set offers another fractal structure with a
nontrivial boundary.
Remarkably, this boundary can be decomposed into elementary domains
that are approximated with high precision by relatively simple
equations.
These domains take the form of cardioids at the cluster centers
(the red curve in Fig.~\ref{fig:Cardioid}) or circles at the
non-root nodes of the tree (green curves), respectively:
\[
c-c_0 = r_0\,e^{i\,\phi}\left(1-\frac{1}{2}\,e^{i\,\phi}\right)
\quad\text{and}\quad
c-c_0 = r_0\,e^{i\,\phi}.
\]
For further details, see~\cite{dolotin2008}, which is dedicated
to this analysis.

\begin{figure*}[t!]
    \centering
    \includegraphics[width=\linewidth]{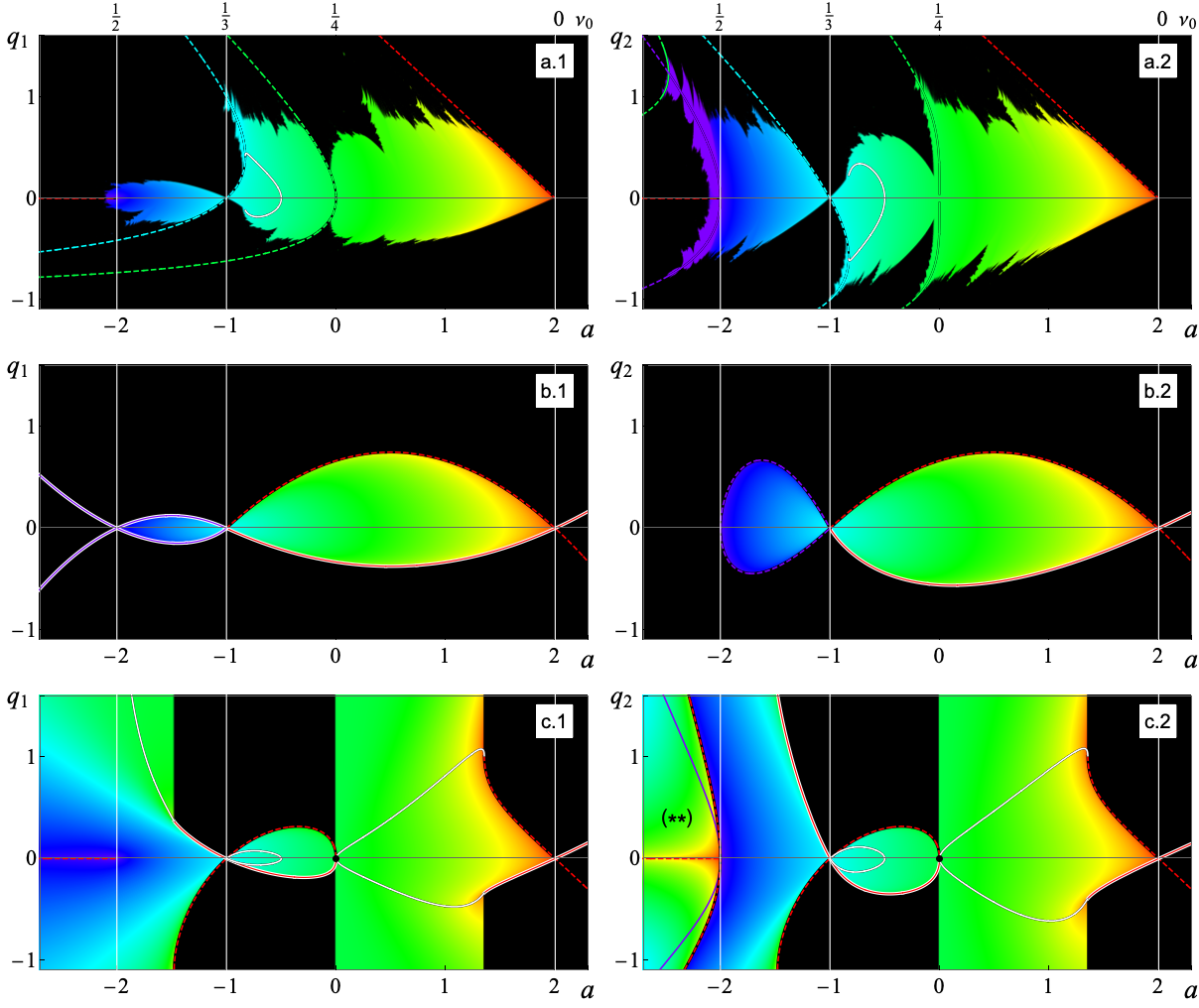}\vspace{-0.2cm}
    \caption{\label{fig:Domain}
    {\bf Approximate domains of the H\'enon set}.
    H\'enon set (a.) compared to its integrable approximations
    using McMillan mappings in the first (b.) and second (c.)
    orders of perturbation theory.
    The bottom two rows show the rotation number evaluated in the
    simply connected regions around the origin and around the
    stable 2-cycle in the area marked with (**) in plot (c.2);
    otherwise, (**) area should be uniformly colored purple,
    indicating mode-locking within two islands.
    Solid and dashed lines represent stable and unstable $n$-cycles,
    colored according to their rotation numbers (red for fixed points,
    purple for 2-, cyan for 3-, and green for 4-cycles).
    The white line represents the coordinates of the twistless orbits.
    For plots (b.) and (c.), the white lines with color show the 
    intersections with homo/heteroclinic connections.
    }
\end{figure*}

Similarly, we aim to explain the shapes of the ``bulbs'' that
form the fractal body of the H\'enon set.
The top row in Fig.~\ref{fig:Domain} illustrates the rotation
number without mode-locking filtering.
Here, unstable and chaotic orbits are rendered in black, making
the shape of the set clearer.

Two distinct types of boundaries emerge to determine stability.
``Ripped'' boundaries, which arise from the overlap of island
chains far from the system's main resonances, define one type
of stability limit.
For the quadratic H\'enon map, such regions include $a\in(0,1)$
or $a\in(-2,-1.5)$.
These boundaries correspond to $\epsilon = 1$ in the circle map
but no longer align with any algebraic curve.

Resonance-driven boundaries, absent in the circle map, are linked
to $\tau_0 = 0$ resonances and their associated singular tongues.
Near resonances, these boundaries follow algebraic curves tied to
unstable $n$-cycles (dashed lines in color).
In integrable systems, these are the only type of boundaries
(if any) in the parameter space that form well-defined bulbs
(see plots b. and c.).
Further from the resonance value of $\nu_0$, in chaotic systems
the stability boundary deviates from the $n$-cycle.

To provide an alternative to $n$-cycle analysis, we now employ
nonlinear integrable approximations.
For the quadratic H\'enon map, perturbation theory can be
used to construct approximate invariants of motion that are
conserved to a specified order of smallness parameter 
$\varepsilon$ (not to be confused with the circle map parameter
$\epsilon$):
\begin{equation}
\label{math:KK'}
\K^{(n)}[p',q'] - \K^{(n)}[p,q] = \mathcal{O}(\varepsilon^{n+1}).
\end{equation}
The invariant can be sought in the form of a polynomial:
\[
\K^{(n)} =
    \K_0 + \varepsilon\,\K_1 + \varepsilon^2\,\K_2 + \ldots +
    \varepsilon^n\,\K_n,
\]
where $\K_m$ consists of homogeneous polynomials in $p$ and $q$
of $(m+2)$ degree, with coefficients determined by satisfying the
perturbation equation Eq.~(\ref{math:KK'}).
Thorough discussion of this perturbation theory, along with
higher-order analyses, will be provided in a subsequent publication.
For now, the first two orders are covered
in~\cite{zolkin2024MCdynamics,zolkin2024MCdynamicsIII}, with a
summary of the essential points provided here.

Interestingly, the first two orders of this theory yield a general
result
\begin{equation}
\label{math:Kapprox}
\begin{array}{l}
\ds \K^{(2)}[p,q] = p^2 - a\,p\,q + q^2
    - \varepsilon\,\frac{b}{a+1}\,(p^2 q + p\,q^2)\,+   \\[0.35cm]
\qquad\ds +\,
    \varepsilon^2\left(\left[
        \frac{b^2}{a\,(a+1)} - \frac{c}{a}
    \right] p^2q^2 \right) + \ldots
\end{array}
\end{equation}
that matches the integrable symmetric McMillan map.
For the quadratic H\'enon map,
\[
f(q) = a\,q+q^2,
\]
the first- and
second-order integrable approximations yield forces:
\[
f_\text{SX-1}(q) =
    \frac{a\,(a+1)+q}{(a+1)-q}\,q =
    a\,q + q^2 + \frac{q^3}{a+1} + \mathcal{O}(q^4),
\]
and
\[
\ds f_\text{SX-2}(q) =
    \frac{a\,(a+1)+q}{(a+1)-q+\frac{1}{a}\,q^2}\,q =
    a\,q + q^2 + \mathcal{O}(q^4).
\]

\newpage
The middle row of Fig.~\ref{fig:Domain} illustrates the stability
region for the SX-1 approximation.
At this order, the motion is always bounded, resulting in two
well-defined bulbs for $\nu_0 < 1/3$ and $\nu_0 > 1/3$.
However, this order is insufficient to accurately match $\tau_0$
for the quadratic H\'enon map and does not account for twistless
orbits at the origin~\cite{zolkin2024MCdynamics}.

In the SX-2 order, $\tau_0$ is exactly
matched~\cite{zolkin2024MCdynamicsIII}.
This approximation reveals a twistless orbit and additionally, it
captures the formation of a bulb for $a\in(-1,0)$, corresponding
to $\nu_0\in(1/4,1/3)$, as well as areas linked to
$\mathcal{T}_{1/2}$ tongues for $a<-1/2$, marked with (**).

However, convergence becomes increasingly challenging in regions
with ripped boundaries, where it is necessary to determine ``the
critical value of $\epsilon$'' rather than relying on specific
$n$-cycle coordinates.
In the integrable approximation, for values of $a$ in the range
$a\in(0,(-1+3\,\sqrt{57})/16)$ and $a<(-1-3\,\sqrt{57})/16$,
unbounded motion around the origin is observed, signaling that
the order of the approximation is insufficient to accurately
capture the dynamics.
Higher-orders reveal additional resonances, such as $1/5,\,2/5$,
and $1/6$, but these models do not yield a well-defined integrable
map, further complicating the analysis of the system's behavior.

\begin{figure*}[t!]
    \centering
    \includegraphics[width=\linewidth]{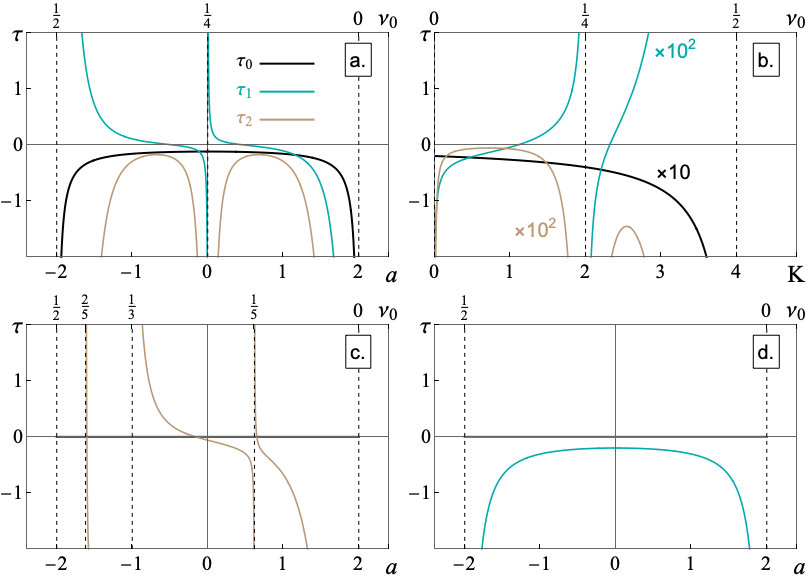}\vspace{-0.1cm}
    \caption{\label{fig:Twist}
    {\bf Twist}. The first three twist coefficients are shown as
    functions of the parameter for various transformations in
    McMillan form:
    (a.) cubic map, $f_+(p) = a\,p+p^3$,
    (b.) Chirikov map, $f(p) = 2\,p+\mathrm{K}\,\sin p$,
    (c.) fourth-power, $f_+(p) = a\,p+p^4$, and
    (d.) fifth-power, $f_+(p) = a\,p+p^5$, mappings.
    The scale at the top indicates the corresponding values of the
    bare rotation number, $\nu_0$.
    In plot (b.), additional scaling (shown in color) is utilized
    to improve the visual representation.
    }
\end{figure*}

\newpage

\section{\label{sec:Results}Results \& Discussion}

In this section, we present bifurcation diagrams for various
mappings and discuss their key features and implications.
By now, we assume the reader is familiar with interpreting these
diagrams, particularly with the aid of the first twist
coefficients.
Figure~\ref{fig:Twist} illustrates $\tau_0$ through $\tau_2$ for\
all mappings discussed.
The scale at the top, labeled with values of the bare rotation
number $\nu_0$, helps in identifying singularities, while the
zero crossings of the lowest nonzero twist coefficient indicate
the presence of twistless orbits.

\subsection{H\'enon sets and sextupole magnet}

A significant application of the quadratic H\'enon map, in its
original form Eq.~(\ref{math:T2Henon}), lies in modeling horizontal
dynamics in accelerator lattices with thin sextupole magnets
(see~\cite{zolkin2024MCdynamics,zolkin2024MCdynamicsIII} for
details).
In this context, the variables $(x,y)$ represent normalized Floquet
coordinates derived from the horizontal position $z$ and its
longitudinal derivative, $\dot{z}=\dd z/\dd s$, where $s$ is the
azimuthal coordinate of the accelerator ring~\cite{SYLee4}.

\begin{figure*}[t!]
    \centering
    \includegraphics[width=\linewidth]{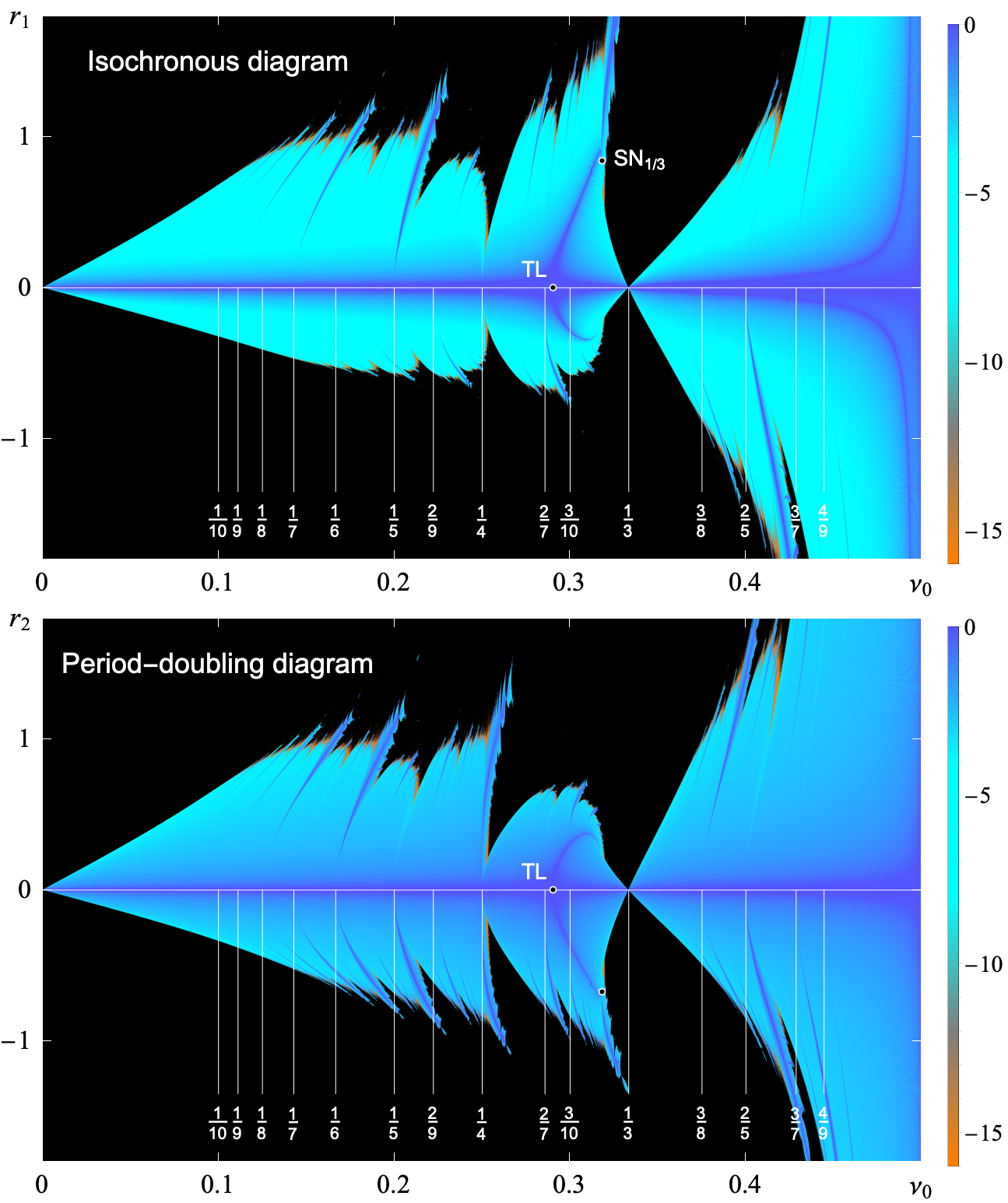}\vspace{0.9cm}
    \caption{\label{fig:Fractal1}
    {\bf H\'enon set}.
    Isochronous and period-doubling diagrams for the H\'enon map
    in its original form.
    The vertical axis represents the radial vector from the origin
    to the corresponding symmetry line, while the horizontal axis
    shows the bare rotation number.
    The generalized alignment index (GALI) is used for coloring.
    }
\end{figure*}

Building on our prior explorations and following H\'enon's intent
we conclude by presenting a final plot in the original form,
Fig.~\ref{fig:Fractal1}, where we employ radius vectors along both
symmetry lines, $r_{1,2} = \sqrt{x_{1,2}^2+y_{1,2}^2}$, instead of
their projections, $x_{1,2}$, and use $\nu_0$ in place of the trace
parameter, $a$.
To reveal the internal structure of the tongues, the GALI indicator
is used for coloring.

This plot holds critical experimental relevance, as it provides
the {\it dynamic aperture}, which quantifies the stability of
particles in the accelerator ring for given parameter settings.
As noted earlier, the variation in scaling near $\nu_0 = 1/2$
compared to Fig.~\ref{fig:FractalNG} is linked to the determinant
of the Jacobian for the transformation~(\ref{math:scaling})
between different forms of the map.
This connection can be further clarified with the insights
provided by Fig.~\ref{fig:dNUdJ}.

\begin{figure*}[t!]
    \centering
    \includegraphics[width=\linewidth]{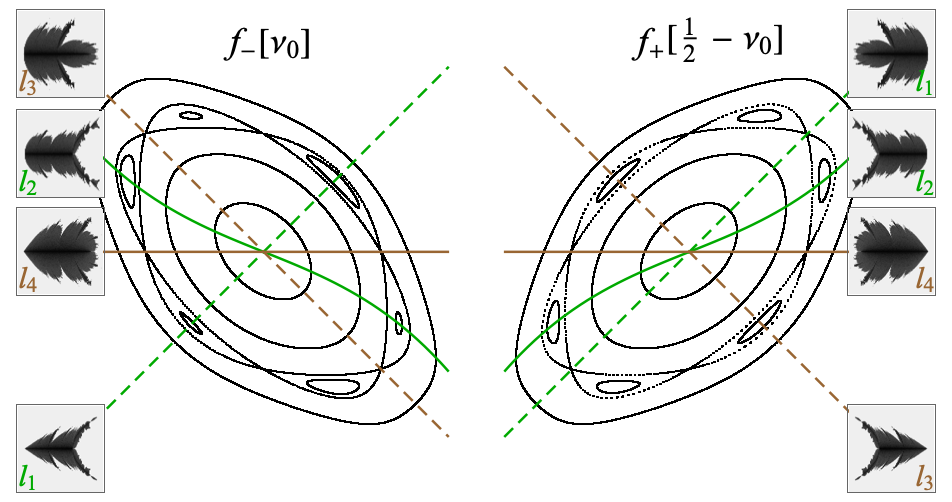}
    \caption{\label{fig:OddSym}
    {\bf Odd force function}.
    This figure illustrates the correspondence between phase space
    portraits and stability diagrams of the cubic mappings
    $f_\pm(p) = a\,p \pm p^3$, demonstrating the symmetry property
    $\nu_0 \leftrightarrow \frac{1}{2}-\nu_0$.
    In this case, the chain of islands is formed by either two
    symmetric stable and two unstable 3-cycles (left) or a pair of
    stable and unstable 6-cycles.
    The first two symmetry lines, $l_{1,2}$, are depicted in green,
    while an additional pair of independent symmetry lines, $l_{3,4}$,
    is shown in brown.
    Stability diagrams for $f_+$ are presented in Fig.~\ref{fig:Bugs}.
    }
\end{figure*}

\subsection{Homogeneous H\'enon mappings}

In this Subsection we consider a family of mappings:
\[
\T_k:\qquad
\begin{array}{l}
x' = a_1\,x + b_1\,y + G_k(x,y),    \\[0.25cm]
y' = a_2\,x + b_2\,y + H_k(x,y),
\end{array}
\]
where $G_k$ and $H_k$ are homogeneous polynomials of degree $k>2$.
Through similar reasoning as for the quadratic case, these mappings
can be transformed into the form (\ref{math:T2Henon}) with a single
intrinsic parameter $\psi$ and a force function $F_\pm(x)=\pm x^k$.
Just like the quadratic H\'enon map, these mappings have physical
interpretations in accelerator physics.
They model rings with higher-order multipole magnets: octupoles
($k=3$), decapoles ($k=4$), and duodecapoles ($k=5$).
The sign reflects the focusing or defocusing nature of the magnet.
While octupoles\cite{octuGareyte1997,octuLeemann2011,octuPlassard2021},
alongside sextupoles, quadrupoles, and dipoles,
are fundamental components of any modern accelerator, higher-order
multipoles are increasingly used for advanced corrections in
machine design~\cite{multOhuchi2022}.

To simplify the equations further, we apply transformation
(\ref{math:HMH}) and an additional isotropic scaling
$(q,p) \rightarrow \varepsilon\,(q,p)$, where
$\varepsilon^{k-1}\,\sin\psi = 1$.
This leads to a McMillan form mapping with
\begin{equation}
\label{math:MHforce}    
    f_\pm(q) = a\,q \pm q^k,
    \qquad
    a = 2\,\cos\psi.
\end{equation}

\subsubsection{Cubic H\'enon map}

As a specific example, we analyze the cubic transformations with
$f_\pm(p) = a\,p \pm p^3$, representing a typical odd force
function.
For mappings in McMillan form that satisfy $f(p)=-f(-p)$, an
additional spatial symmetry arises.
Specifically, these mappings commute with the area-preserving
involution:
\[
    \T = \Rt(\pi)\circ\T\circ\Rt^{-1}(\pi).
\]
This property generates a distinct class of transformations:
\[
\mathrm{Q}_1 = \Rf(\pi/4)\circ\Rt(\pi) = \Rf(3\,\pi/4):\qquad
\begin{array}{l}
    q'=-p,  \\[0.25cm]
    p'=-q,
\end{array}
\]
and
\[
\mathrm{Q}_2 = \T\circ\mathrm{Q}_1:\qquad
\begin{array}{l}
    q'=-q,  \\[0.25cm]
    p'= p - f(q),
\end{array}
\]
which constitute an independent symmetry group, distinct from
the one generated by $\R_1=\Rf(\pi/4)$~\cite{roberts1992revers}.
As a result, the system exhibits two additional symmetry lines
\begin{equation}
\label{math:4syms}
\begin{array}{ll}
    l_1:\,p=q,   & \qquad l_2:\,p=f(q)/2,  \\[0.25cm]
    l_3:\,p=-q,   & \qquad l_4:\,p=0.
\end{array}
\end{equation}

\begin{figure*}[t!]
    \centering
    \includegraphics[width=\linewidth]{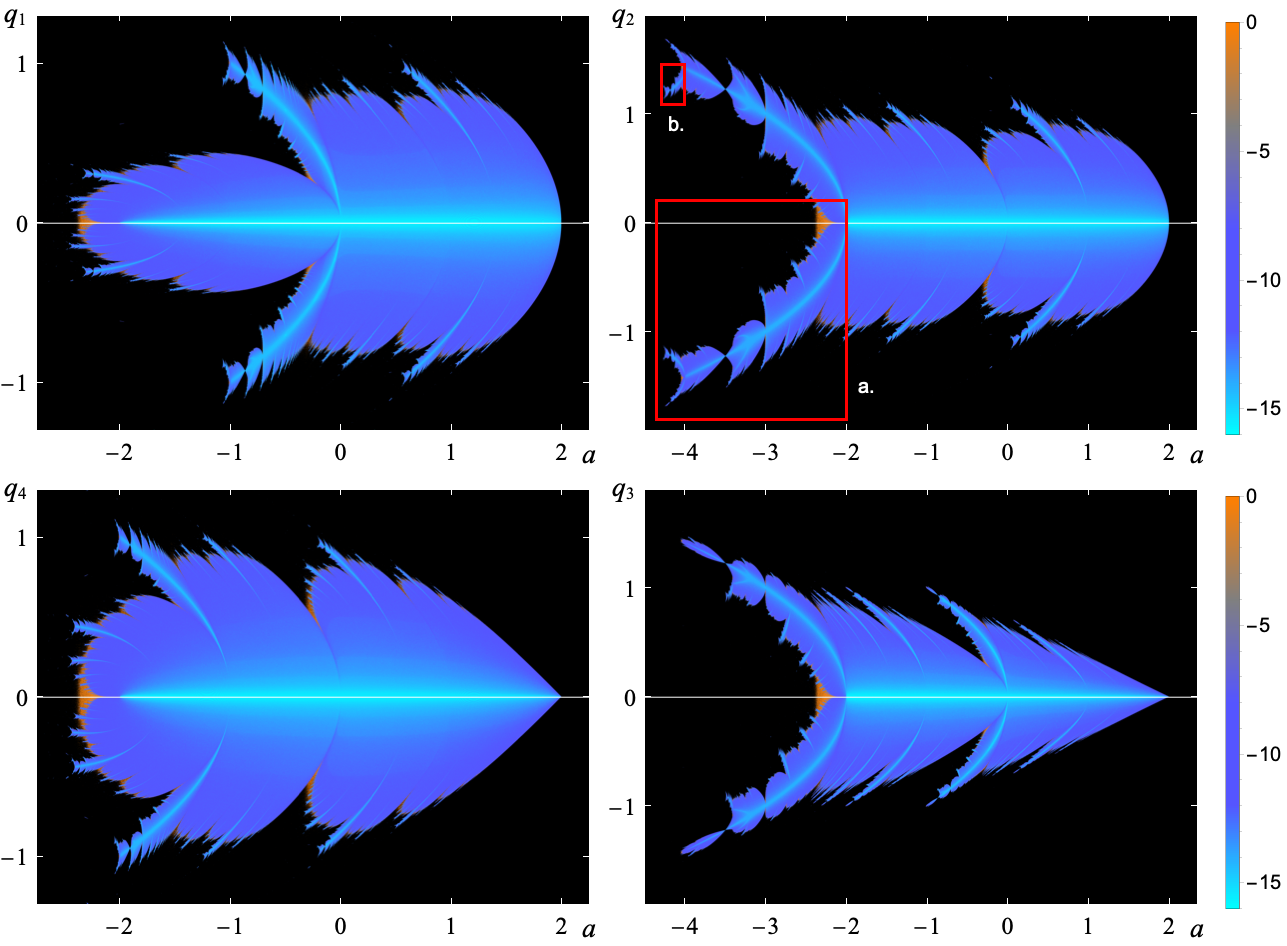}
    \caption{\label{fig:Bugs}
    {\bf Beetles and Bugs}.
    Stability diagrams (REM) for the homogeneous cubic map in McMillan 
    form, $f_+(p) = a\,p + p^3$.
    The vertical axes represent the $q$ coordinate of points along
    the symmetry lines $l_{1,2,3,4}$.
    Fig.~\ref{fig:BugsZOOM} compares areas (a.) and (b.).
    }
\end{figure*}

\newpage

While the presence of four distinct fractals adds some complexity
to the interpretation, this is offset by the fact that the
additional spatial symmetry simplifies the construction of
stability diagrams (as the lower half $q<0$ of each fractal can
be inferred using mirror symmetry), and that it suffices to
analyze $f_+$, see Fig.~\ref{fig:Bugs}.
The fractals for $f_-$ can be reconstructed by simply reversing
the direction of $a$ and swapping the indices of $l_1$ and $l_3$.
An illustrative example is provided in Fig.~\ref{fig:OddSym}.
Additionally, when plotting rotation numbers, the color values must
be transformed as:
$
    \nu \rightarrow \frac{1}{2} - \nu,
$
which can be readily understood as a consequence of Danilov theorem
\cite{nagaitsev2020betatron}.

\subsubsection{Fourth and fifth power functions}

\begin{figure*}[p!]
    \centering
    \includegraphics[width=\linewidth]{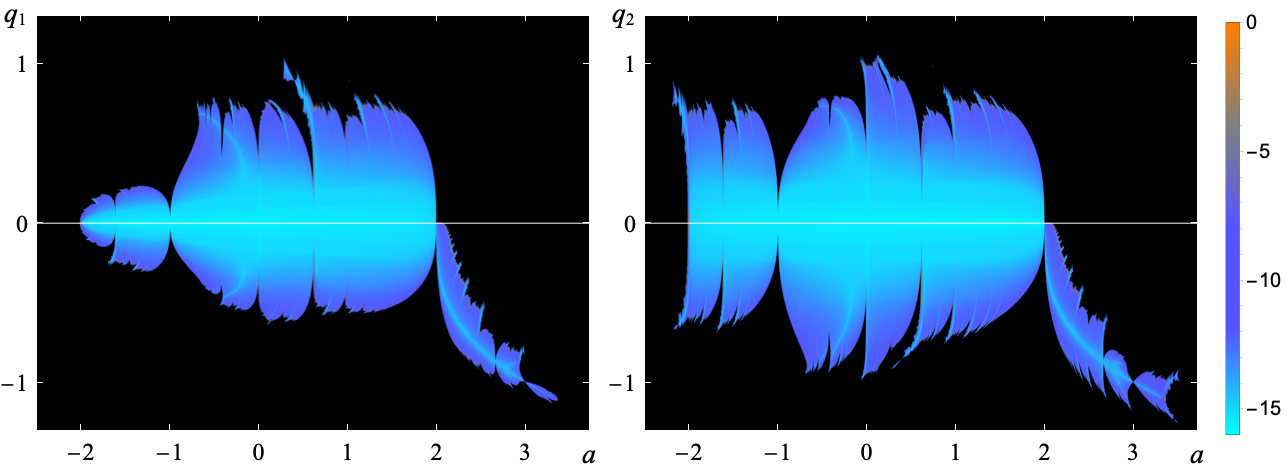}
    \caption{\label{fig:Elephant}
    {\bf Elephant}.
    Stability diagrams (REM) for the homogeneous fourth-power map
    in McMillan  form, $f_+(p) = a\,p + p^4$.
    The vertical axes represent the $q$ coordinate of points along
    the symmetry lines $l_{1,2}$.}
\end{figure*}

\begin{figure*}[p!]
    \centering
    \includegraphics[width=\linewidth]{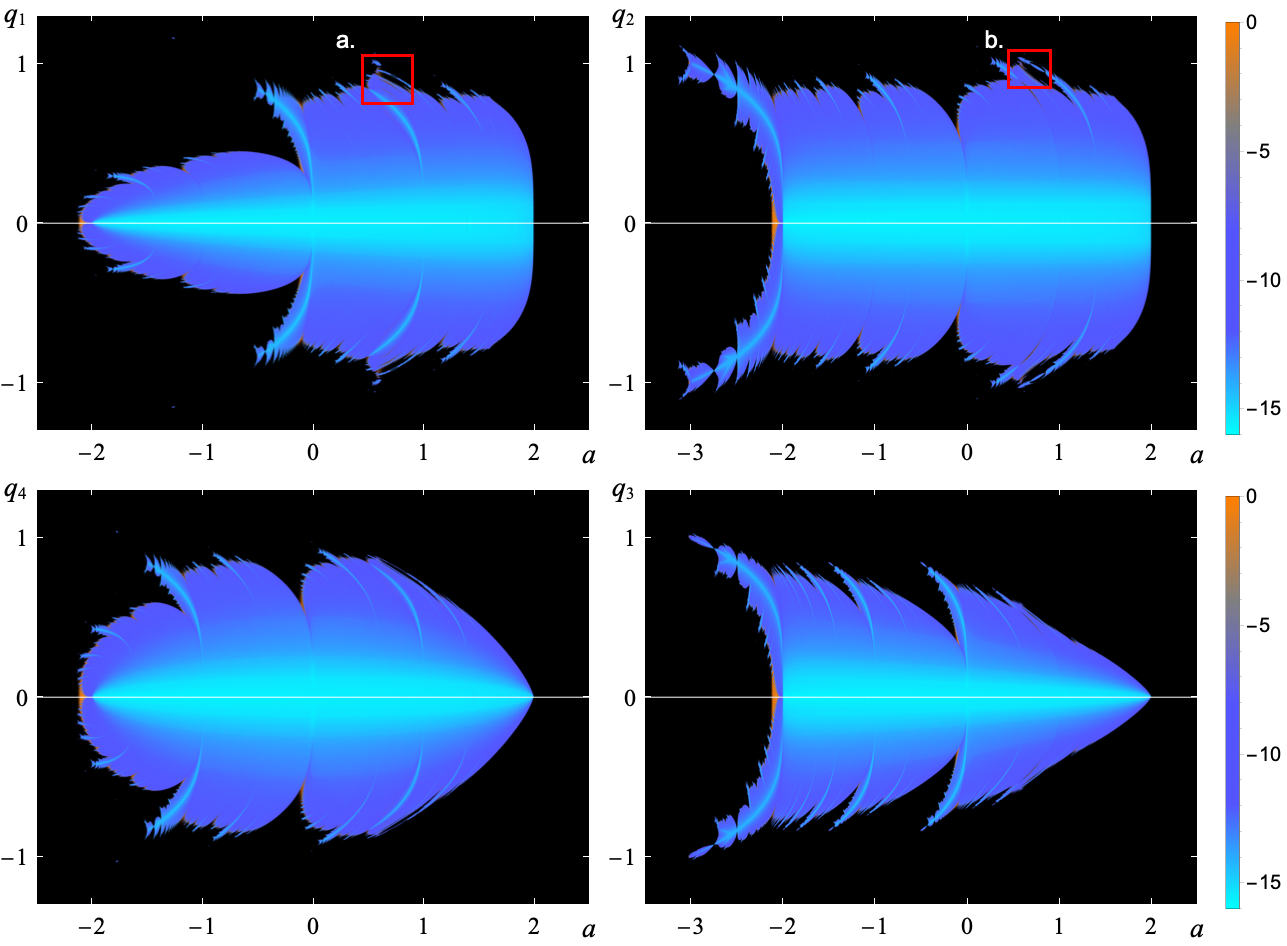}
    \caption{\label{fig:Hyppo}
    {\bf Hippo}.
    Stability diagrams (REM) for the homogeneous fifth-power map
    in McMillan  form, $f_+(p) = a\,p + p^5$.
    The vertical axes represent the $q$ coordinate of points along
    the symmetry lines $l_{1,2,3,4}$.}
\end{figure*}

Next, we present results for two additional mappings, characterized
by the force functions
\[
    f(p) = a\,p+p^4
    \qquad\text{and}\qquad
    f(p) = a\,p+p^5
    \]
as shown in Figs.~\ref{fig:Elephant} and \ref{fig:Hyppo},
respectively.
These systems exhibit an atypical behavior, as they lack both
lower-order terms $q^2$ and $q^3$.
Consequently, the usual integrable approximation cannot be applied,
and the first twist coefficient, $\tau_0$, vanishes for all
$a\in(0,2)$.
To gain insights into the slopes of resonance tongues and twistless
orbits, it is necessary to consider $\tau_{1,2}$.

For instance, in plot (c.) of Fig.~\ref{fig:Twist}, we observe
that the resonances at $\nu_0 = \frac{1}{3},\,\frac{1}{5}$, and
$\frac{2}{5}$ are singular, along with the emergence of three
twistless bifurcations at the origin.
Notably, only one associated anti-tongue is visible in
Fig.~\ref{fig:Elephant}, as the other two are too narrow in the
parameter space to be easily discerned.
Even this visible anti-tongue is barely distinguishable near the
origin due to the weak dependence of amplitude on the rotation
number.
For $a>2$ in Fig.~\ref{fig:Elephant}, the ``trunk of an elephant''
formation arises, representing the integer tongue $\mathcal{T}_0$
associated with the second fixed point
\[
\z^{(1-2)} = \sqrt[3]{2-a}\times(1,1),
\]
becoming stable as the origin loses stability at $a=2$.

\begin{figure*}[p!]
    \centering
    \includegraphics[width=\linewidth]{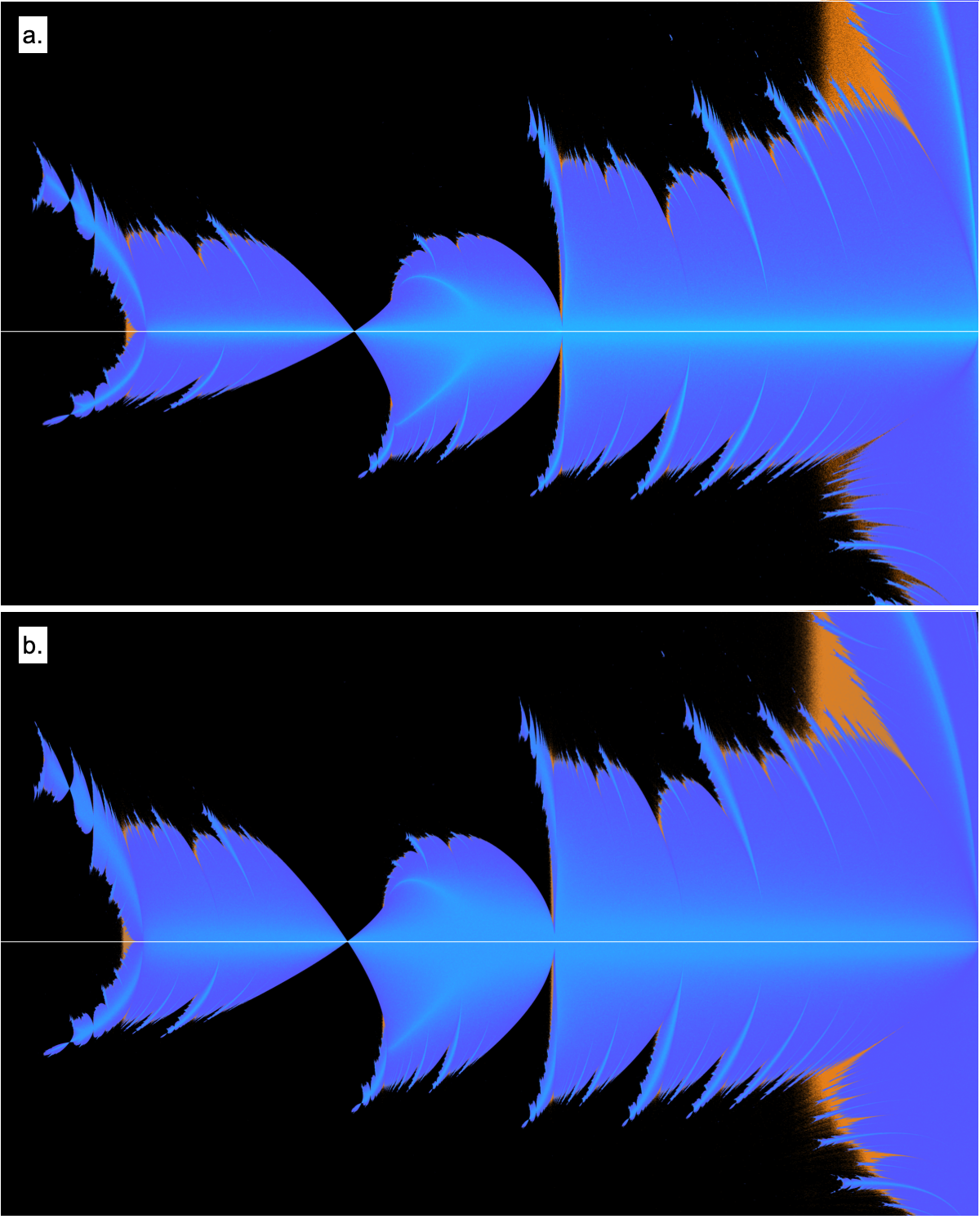}
    \caption{\label{fig:BugsZOOM}
    {\bf Self-similarity}.
    Magnified views of the regions highlighted by red rectangles
    in Fig.~\ref{fig:Bugs}, showing self-similar structures after
    subtracting the corresponding coordinates of the 2-cycle (a.)
    or 4-cycle (b.).
    }
\end{figure*}

\begin{figure*}[p!]
    \centering
    \includegraphics[width=\linewidth]{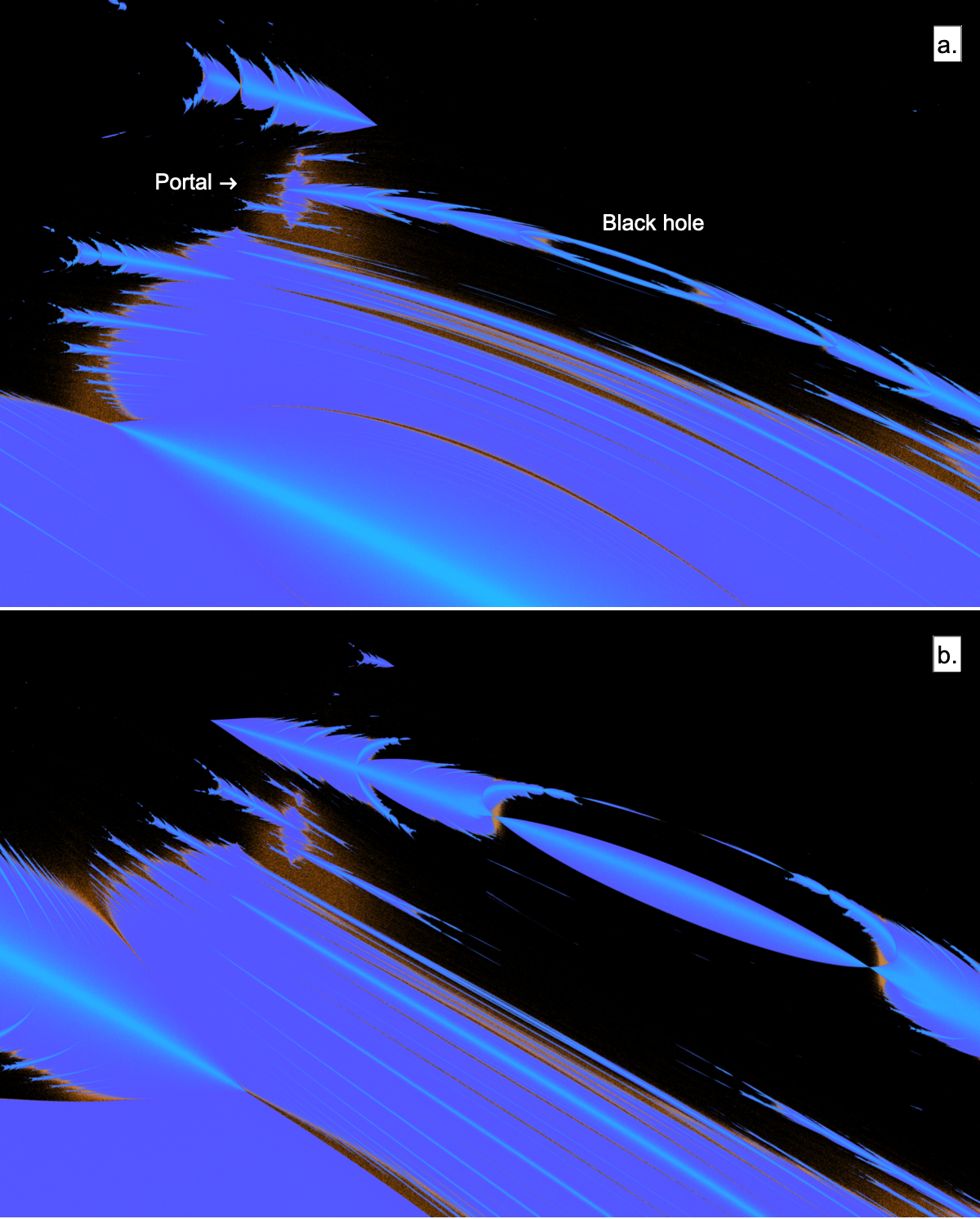}
    \caption{\label{fig:Portal}
    {\bf Portal and Black hole}.
    Magnified views of the areas highlighted with red rectangles
    in Fig.~\ref{fig:Hyppo}.
    }
\end{figure*}

\newpage
In the fifth-power map, as with the cubic map, the first non-zero
twist coefficient never vanishes.
This implies the absence of twistless bifurcations at the origin;
however, such bifurcations do occur within resonance tongues and
are associated with higher-order ($n>1$) periodic orbits.
Once again, four symmetry lines are required, and the overall
fractal structure closely resembles that in Fig.~\ref{fig:Bugs}.

At large amplitudes, the tongues exhibit highly complex overlap
structures and bifurcations.
To illustrate this, we provide magnified views of two regions
from the top plots in Fig.~\ref{fig:Hyppo}, which might vaguely
resemble the head of a hippopotamus.
These magnifications are presented in panels (a.) and (b.) of
Fig.~\ref{fig:Portal}.

While a detailed analysis of these systems is beyond the scope of
this discussion, a few noteworthy observations can be made.
In Fig.~\ref{fig:Portal}, we observe that distinct tongues can
reconnect (as seen in the region labeled ``black hole''), form
intricate clusters (``portals'').
Additionally, ``splits'' in the main body of a fractal that are
filled wit chaotic trajectories (ca be seen in lower parts of both
plots) undergo metamorphoses into higher-order tongues.

\subsection{Chirikov map}

Next, we revisit the renowned Chirikov standard
map~\cite{chirikov1969research,chirikov1979universal}, sometimes
referred to as the Taylor-Greene-Chirikov map:
\begin{equation}
\label{math:ChirMap}    
\T_\mathrm{Chirikov}:\qquad
\begin{array}{l}
I'      = I + F(\theta),    \\[0.25cm]
\theta' = \theta + I',
\end{array}
\end{equation}
where $F(\theta) = \mathrm{K}\,\sin\theta$.
Due to the periodicity of $F$, $\T_\mathrm{Chirikov}$ can be
interpreted in three distinct settings:
the plane $(\theta,I)\in\mathbb{R}^2$,
a cylinder by taking $\theta\mod 2\pi$,
or a torus by applying $\mod 2\pi$ to both equations.
For additional details, refer to~\cite{ZKN2024PolII}.

This map extends the circle map (\ref{math:CircleMap}), which is
obtained by setting $I' = \Omega$ and $\mathrm{K} = \epsilon$ in
the second equation.
It also has direct applications in accelerator physics,
particularly in describing longitudinal dynamics in machines with
thin RF-stations, likely one of B. Chirikov’s original motivations.

Using the change of variables:
\begin{equation}
\label{math:CMC}
\begin{array}{ll}
    \theta = p,         &\qquad
    q = \theta - I,     \\[0.25cm]
    I  = p - q,         &\qquad
    p = \theta,
\end{array}
\end{equation}
the Chirikov map can be expressed in McMillan form with
\[
    f(p) = 2\,p + F(p),
\]
and the trace of the transformation given by:
\[
\Tr\,\dd\T_\mathrm{Chirikov} = 2 + \mathrm{K}\,\cos p.
\]
The map has two fixed points, $(0,0)$ and $(\pi,\pi)$, which are
stable under the respective conditions
\[
-4 < \mathrm{K} < 0
\qquad\text{and}\qquad
0 < \mathrm{K} < 4.
\]

\begin{figure}[t!]
    \centering
    \includegraphics[width=\linewidth]{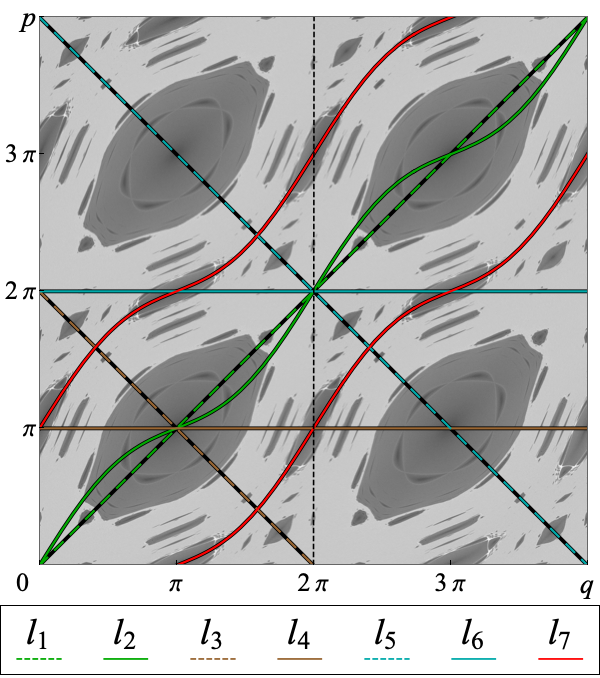}
    \caption{\label{fig:ChirikovPS}
    {\bf Symmetries of Chirikov map}.
    Phase space diagram of the Chirikov map, showcasing the
    associated symmetry lines, color-coded as indicated in the
    legend.
    }
\end{figure}

The force function is odd with respect to both fixed points, as
the map commutes not only with $\Rt(\pi)$, but also with a more
general involution:
\[
\begin{array}{l}
    q'=-q + 2\,\pi\,k,  \\[0.25cm]
    p'=-p + 2\,\pi\,k,
\end{array}
\qquad
k\in\mathrm{Z}.
\]
As a result, the map exhibits six symmetry lines,
see Fig.~\ref{fig:ChirikovPS}:
\[
\begin{array}{ll}
    l_1:\,p=q,          & \qquad l_2:\,p=f(q)/2,    \\[0.25cm]
    l_3:\,p=2\,\pi-q,    & \qquad l_4:\,p=\pi,       \\[0.25cm]
    l_5:\,p=-q,         & \qquad l_6:\,p=0.
\end{array}
\]
Here, $l_3$ and $l_5$ are not independent, so the arithmetic
quasiperiodicity of $f$ introduces only one additional symmetry
line, compared to a generic odd force.

The map also commutes with a shift along the main diagonal $p=q$:
\[
\mathrm{S}:\qquad
\begin{array}{l}
    q'= q + 2\,\pi\,k,  \\[0.25cm]
    p'= p + 2\,\pi\,k,
\end{array}
\qquad
k\in\mathrm{Z}.
\]
This shift, though not an involution, represents an example of weak
symmetry.
The associated transformations $\mathrm{L}_1 = \R_1\circ\mathrm{S}$
and $\mathrm{L}_2 = \T\circ\mathrm{L}_1$, when considered
$\mathrm{mod}\,2\,\pi$, introduce symmetry lines:
\[
    p = q + 2\,\pi\,k\,
    \qquad\mathrm{and}\qquad
    l_{7,8}:\,p = \frac{f(q)}{2} \pm \pi.
\]

\begin{figure*}[p!]
    \centering
    \includegraphics[width=\linewidth]{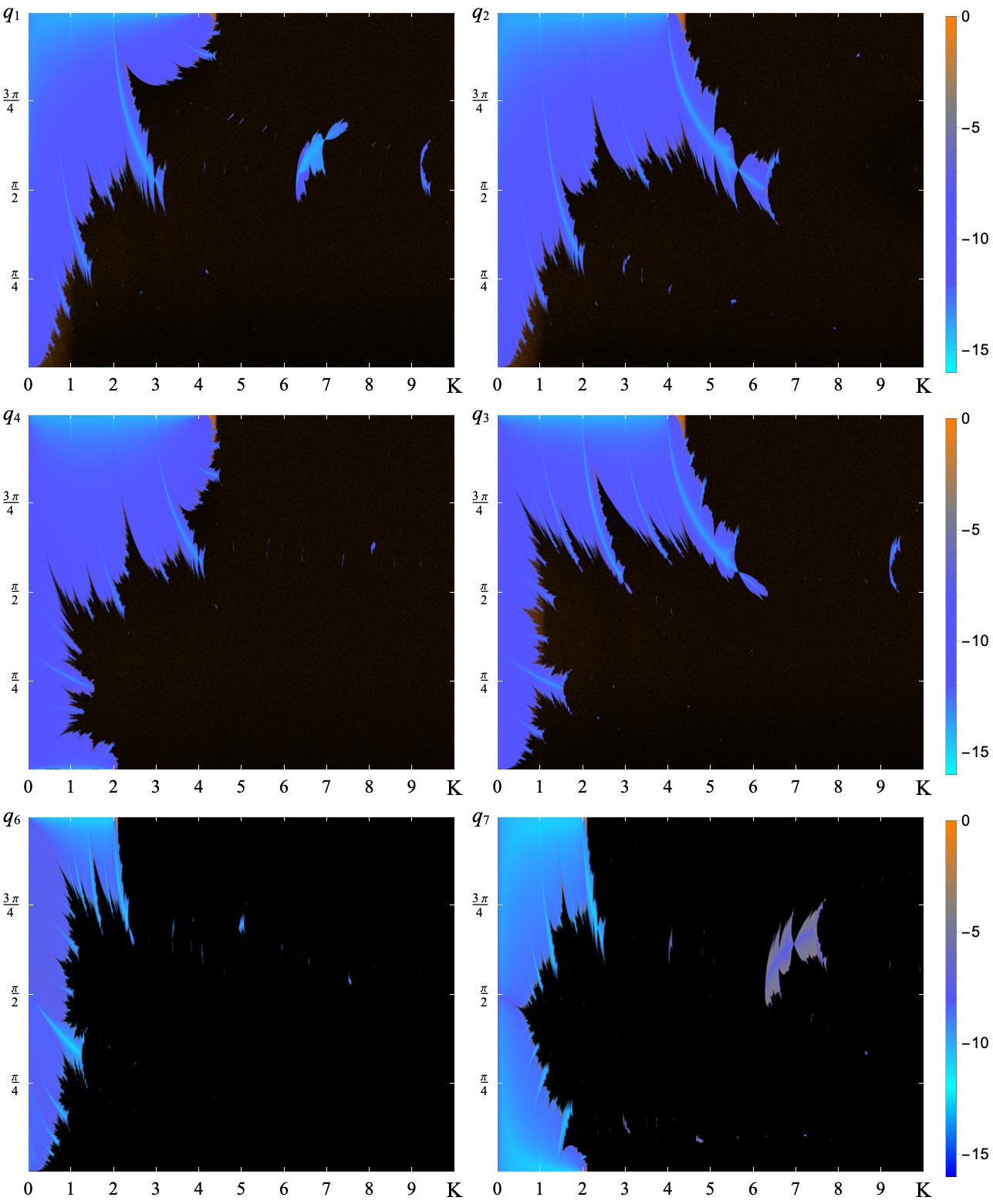}
    \caption{\label{fig:Chirikov}
    {\bf Chirikov map}.
    Stability diagrams (REM) for the Chirikov map in McMillan form,
    $f(p) = 2\,p + \mathrm{K}\,\sin p$.
    The vertical axes indicate the $q$-coordinates of points along
    the symmetry lines $l_{1,2,3,4,6,7}$, see
    Fig.~\ref{fig:ChirikovPS} for reference.
    }
\end{figure*}

\newpage
Restricting to $\mathrm{K}>0$ (without loss of generality),
only the fixed point at $(\pi,\pi)$ can be stable, with its
rotation number given by:
\[
    \nu_0 = \arccos(1-\mathrm{K}/2)/(2\,\pi).
\]
Figure~\ref{fig:Chirikov} shows the stability diagrams along
the symmetry lines $l_{1,2,3,4,6,7}$, providing a comprehensive
visualization of the map’s dynamical features.
This concludes our exploration.

\vspace{-0.2cm}
\subsection{Fracture of twistless structure}

In addition to the diagrams in Fig.~\ref{fig:Bugs}, the
supplementary materials include three types of animations that
explore additional aspects and raise intriguing questions.
The playlist of animations, along with the corresponding code,
is also accessible on
\href{https://www.youtube.com/playlist?list=PLZUmQBEOKX4JQFZFBS8Pm66ECZNlzGCeM}{YouTube} and
\href{https://github.com/FractalTongues/fractal}{GitHub}.

First, the videos labeled \texttt{*\_scan.mp4} demonstrate the
correspondence between phase space portraits and stability diagrams,
offering a dynamic perspective.
Second, the animations titled \texttt{*\_rotation.mp4} showcase
stability diagrams generated by rotating the cross-section $p=q$
instead of using a ``true'' symmetry line.

Finally, the most compelling animations, labeled \texttt{*\_mix.mp4},
investigate the interplay of various parameter ratios, specifically 
$c/b$ and $c/d$.
These animations reveal the evolution of twistless structures as the
parameter $c$ is varied.
For instance, when $b=0$, no twistless bifurcation occurs at the
origin, but for the quadratic H\'enon map $c=0$, the twistless
structure appears as a unified entity in parameter space.
However, as $c$ increases, the twistless structures evolve,
eventually crossing major resonances and fracturing into multiple
smaller segments.
A detailed investigation into how the coherence of such structures
is disrupted by various perturbations warrants a separate
publication.
As previously mentioned, while twistless orbits can stabilize
dynamics at mid-range amplitudes, their influence can be
overridden by the intrusion of higher-order tongues.

\vspace{-0.1cm}
\section{\label{sec:Summary}Summary}

This study explores the visualization of dynamics in reversible
symplectic mappings of the plane, focusing on the challenges posed
by reducing the combined space of variables and parameters.

A dynamical system is reversible if there is an involution in phase
space which reverses the direction of time.
For discrete-time systems, this means the transformation is a
composition of two distinct involutions, forming a group and
defining two families of symmetries.
Moreover, we demonstrate that groups of symmetric orbits arising
from typical bifurcations of a fixed point, such as saddle-node,
transcritical, period-doubling, $n$-island chain, and touch-and-go,
can be identified along at least one of the two principal symmetry
lines, addressing questions first posed in H\'enon's original work.
To provide a more comprehensive description of these systems, the
paper introduces the use of both isochronous and period-doubling
diagrams, with the possibility of extending this framework in cases
of multiple reversibility.

Modern chaos indicators, such as the Reversibility Error Method
(REM) and the Generalized Alignment Index (GALI), are employed to
analyze the dynamics further.
These methods prove effective not only in identifying singular and
mode-locked structures (Arnold tongues), but also in resolving
twistless orbits and their associated bifurcations.
Unlike isolated periodic $n$-cycles, which are governed by constant
rotation numbers, twistless orbits follow a different intrinsic
variable: zero twist, defined as the derivative of the rotation
number with respect to the action variable.
These orbits can represent quasiperiodic trajectories that densely
fill closed curves in phase space.

This study provides both qualitative and quantitative analyses of
fractal stability domains in the mixed parameter-symmetry line space,
including approximations based on integrable McMillan maps.
These methods are applied to practical problems, such as visualizing
dynamic aperture in accelerator physics, which underscores their
real-world relevance.
For case studies, we analyzed systems representing longitudinal and
transverse dynamics in accelerator lattices with various types of
thin magnets and RF-station.
We anticipate that a similar approach could be extended to other
fields, such as addressing plasma stability in Tokamaks
(see~\cite{chang2024role}).

This research opens several intriguing avenues for further
investigation.
One compelling question concerns the existence of one-dimensional
structures in phase space, analogous to symmetry lines, that
intersect all asymmetric isolated $n$-cycles.
Another is the potential universality and self-similarity of the
observed diagrams.
For instance, the H\'enon-Heiles system --- a model of planar
stellar motion around a galactic center ---  upon phase space
reduction yields a 2D mapping that exhibits striking structural
similarities to the H\'enon quadratic map~\cite{barrio2020}.
Finally, as highlighted in the preceding section, the detailed
dynamics of twistless orbits during interactions with primary
resonances remain an open topic, presenting fertile ground for
future exploration.

\vspace{-0.1cm}
\section{Acknowledgments}

The authors would like to thank Taylor Nchako (Northwestern
University) for carefully reading this manuscript and for her
helpful comments.
This manuscript has been authored by Fermi Research Alliance, LLC
under Contract No. DE-AC02-07CH11359 with the U.S. Department of
Energy, Office of Science, Office of High Energy Physics.
Work supported by the U.S. Department of Energy, Office of Science,
Office of Nuclear Physics under contract DE-AC05-06OR23177.
I.M. acknowledges that his work was partially supported by the
Ministry of Science and Higher Education of the Russian Federation
(project FWUR-2024-0041).

\newpage
%


\end{document}